%% file: main.tex
\documentclass[acmsmall,screen]{acmart}
\AtBeginDocument{%
  }

\setcopyright{cc}
\setcctype{by}
\acmDOI{10.1145/3808177}
\acmYear{2026}
\acmJournal{PACMSE}
\acmVolume{3}
\acmNumber{FSE}
\acmArticle{FSE170}
\acmMonth{7}
\acmSubmissionID{fse26mainb-p1401-p}
\received{2025-09-03}
\received[accepted]{2026-03-24}

\newcommand{\rv}[1]{{#1}}

\input{packages}

\begin{document}

\title{ReGA: Model-Based Safeguard for LLMs via Representation-Guided Abstraction}

\author{Zeming Wei}
\orcid{0009-0008-2953-0749}
\affiliation{%
  \institution{Peking University}
  \city{Beijing}
  \country{China}
}
\email{weizeming@stu.pku.edu.cn}

\author{Chengcan Wu}
\orcid{0009-0007-9193-0682}
\affiliation{%
  \institution{Peking University}
  \city{Beijing}
  \country{China}
}
\email{2300010746@stu.pku.edu.cn}

\author{Meng Sun}
\orcid{0000-0001-6550-7396}
\affiliation{%
  \institution{Peking University}
  \city{Beijing}
  \country{China}
}
\email{sunmeng@math.pku.edu.cn}
\authornote{Meng Sun is the corresponding author.}

\input{content/0_abstract}



\maketitle

\input{content/1_introduction}

\input{content/2_background}

\input{content/3_approach}

\input{content/4_experiment}

\input{content/5_related}

\input{content/conclusion}

\section*{Data Availability}

Our code is available at \url{https://github.com/weizeming/ReGA}.

\section*{Acknowledgement}
This research was supported by National Natural Science Foundation of China (Grant No. 92582102, 62572013, 62172019) and Beijing Natural Science Foundation, China (Grant No. QY24035).

\bibliographystyle{ACM-Reference-Format}
\bibliography{reference}

\end{document}

%% file: packages.tex
\usepackage[utf8]{inputenc} 
\usepackage[T1]{fontenc}    
\usepackage{url}            
\usepackage{booktabs}       
\usepackage{amsfonts}       
\usepackage{nicefrac}       
\usepackage{microtype}      
\usepackage{xcolor}         

\usepackage{hyperref}
\hypersetup{
    breaklinks,
    colorlinks,
    citecolor=blue
}

\usepackage{microtype}
\usepackage{graphicx}
\usepackage{tabularx}
\usepackage{booktabs} 
\usepackage[utf8]{inputenc} 
\usepackage{url}
\usepackage{booktabs}    
\usepackage{amsfonts}     
\usepackage{nicefrac}       
\usepackage{microtype}     
\usepackage{utfsym}
\usepackage{float}
\usepackage{color}
\usepackage[most]{tcolorbox}
\usepackage[ruled,vlined,linesnumbered]{algorithm2e}
\usepackage{amsmath}
\usepackage{amsthm}
\usepackage{multirow}
\usepackage{multicol}
\usepackage{algpseudocode}
\usepackage{amssymb}
\usepackage{longtable}
\usepackage{soul}
\usepackage{subcaption}
\usepackage{natbib}
\usepackage{bbding}
\usepackage{indentfirst}
\usepackage{cleveref}
\usepackage{pifont}
\usepackage{diagbox}
\usepackage{graphicx}
\usepackage{amsmath,amssymb,amsfonts}
\usepackage{textcomp}
\usepackage{xcolor}
\usepackage{makecell} 

\newcommand{\boxmargin}{5pt}
\definecolor{mynewcolor}{RGB}{235,245,255}
\definecolor{myleftcolor}{RGB}{223,223,223}
\newtcolorbox{myboxc}{
    colback=mynewcolor, 
    colframe=myleftcolor, 
    arc = 0pt, outer arc = 0pt,
    boxsep=0pt, left = 3pt, right = 0pt, top = 0pt, bottom = 0pt, 
    leftrule=3pt, bottomrule=0pt, toprule=0pt, rightrule=0pt,
    left = \boxmargin, right = \boxmargin, top = \boxmargin, bottom = \boxmargin
}

\definecolor{myrqcolor}{RGB}{255,240,240}
\newtcolorbox{myrqbox}{
    colback=myrqcolor, 
    colframe=myleftcolor, 
    arc = 0pt, outer arc = 0pt,
    boxsep=0pt, left = 3pt, right = 0pt, top = 0pt, bottom = 0pt, 
    leftrule=3pt, bottomrule=0pt, toprule=0pt, rightrule=0pt,
    left = \boxmargin, right = \boxmargin, top = \boxmargin, bottom = \boxmargin
}

\newcommand{\answer}[1]{
\begin{center}
\begin{myboxc}
#1
\end{myboxc}
\end{center}
} 

\newcommand{\myrq}[1]{
\begin{center}
\begin{myrqbox}
#1
\end{myrqbox}
\end{center}
}

\setcitestyle{numbers,square,comma}

\newtheorem{definition}{Definition}

%% file: content/0_abstract.tex
\begin{abstract}
Large Language Models (LLMs) have achieved tremendous success in various tasks, yet concerns about their safety and security have emerged. In particular, they pose risks of generating harmful content and are vulnerable to jailbreaking attacks, creating unaddressed security issues regarding their deployments. In the context of software engineering for artificial intelligence (SE4AI) techniques, model-based analysis has demonstrated notable potential for analyzing and monitoring machine learning models, particularly in stateful deep neural networks. However, it suffers from scalability issues when extended to LLMs due to their vast feature spaces. In this paper, we aim to address the scalability issue of model-based analysis techniques for safeguarding LLM-scale models. Motivated by the recent discovery of low-dimensional safety-critical representations that emerged in LLMs, we propose ReGA, a model-based analysis framework with Representation-Guided Abstraction, to safeguard LLMs against harmful prompts and generations. By leveraging safety-critical representations, which are key directions in hidden states that indicate safety-related concepts, ReGA effectively narrows the scalability gap when developing the abstract model for safety modeling. Our comprehensive evaluation shows that ReGA performs sufficiently well in distinguishing between safe and harmful inputs, achieving an AUROC of 0.975 at the prompt level and 0.985 at the conversation level. Additionally, ReGA exhibits robustness to real-world attacks and generalization across different safety perspectives, outperforming existing safeguard paradigms in terms of interpretability and scalability. Overall, ReGA serves as an efficient and scalable solution to enhance LLM safety by integrating representation engineering with model-based abstraction, paving the way for new paradigms to utilize software insights for AI safety. Our code is available at \url{https://github.com/weizeming/ReGA}.
\end{abstract}

%% file: content/1_introduction.tex
\section{Introduction}
\label{sec:intro}

In the past few years, Large Language Models (LLMs) have achieved remarkable success across various tasks, significantly transforming the paradigms of machine learning (ML). By leveraging knowledge from extensive data sets and utilizing advanced next-token decoding strategies, LLMs have made significant advancements in areas such as chat completion~\cite{openai2024gpt4,korbak2023pretraining}, mathematical reasoning~\cite{imani2023mathprompter,ahn2024large}, code generation~\cite{guo2024deepseek,coignion2024performance}, and program repair~\cite{bouzenia2024repairagent,jin2023inferfix} tasks. These accomplishments established the foundational role of LLMs in modern ML and software systems.

However, like traditional ML models and software, current LLMs face significant trustworthiness issues related to their safety and security~\cite{anwar2024foundational,zhang2024fusion,yao2024survey,chen2023combating}. While their advanced reasoning and generation capabilities have facilitated numerous applications, concerns about their potential to produce harmful or toxic content have also emerged. Various evaluations on LLMs have shown that LLMs may positively respond to unsafe requests from users that violate their safety or ethical guidelines~\cite{mazeika2024harmbench,wang2023decodingtrust,sun2024trustllm,zhang2024multitrust}. Furthermore, recent attack techniques have been developed to deceive LLMs into generating harmful content, even if the LLMs successfully reject the request prompt in its original form. These tactics are known as \emph{jailbreaking} attacks~\cite{shen2023do,zou2023universal,wei2023jailbroken,liu2023jailbreaking}, raising further security concerns about LLMs. To prevent confusion, this work refers to the \textit{security risk} as the generation of harmful content by LLMs.

To mitigate such concerns, numerous efforts have been dedicated to aligning LLMs with human values during their pre-training or post-training phases~\cite{korbak2023pretraining,bai2022constitutional,dai2024safe}. Nevertheless, safeguarding LLMs against unsafe prompts and outputs in an effective and efficient manner remains a challenging problem~\cite{anwar2024foundational}. Notably, \textbf{model-based analysis} techniques have demonstrated strong potential in analyzing, inspecting, and repairing Deep Neural Networks (DNNs) through a thread of recent research~\cite{du2019deepstellar,xie2021rnnrepair,du2020marble,ren2023deeparc,xie2023mosaic,qi2023archrepair,wei2024weighted}. For example, DeepStellar~\cite{du2019deepstellar} constructs an abstract model to detect adversarial inputs of recurrent neural networks (RNNs), and Marble~\cite{du2020marble} analyzes the robustness of stateful DNNs through similar model abstraction. By constructing more formal and interpretable abstract models that simulate specific properties of the target DNN with internal hidden-state feature extraction, these works successfully achieve various analysis goals related to the DNN. So far, model-based analysis has become a representative paradigm of leveraging software techniques into DNN analysis~\cite{huang2020survey,martinez2022software}, contributing significantly to the success of \textbf{software engineering for artificial intelligence (SE4AI)} research.

However, existing abstract model extraction-based techniques suffer from scalability hurdles when extending to LLM-scale models, limiting their effectiveness to LLMs with architectures and parameters much larger than those of traditional DNN models. In particular, a recent preliminary work, LUNA~\cite{song2024luna}, explores universal analysis on LLMs through extracting abstract models like Discrete Time Markov Chain (DTMC), but requires a vast amount of abstract states to be effective (ranging from hundreds to millions), limiting the efficiency and effectiveness of the abstract model for analysis. The fundamental bottleneck along this thread is that the feature space of LLMs is too large to be universally modeled by abstract models for different perspectives on trustworthiness. In this context, we refer to the \textit{scalability} issue of model-based analysis as the effectiveness of their modeling when assessing the target property (\textit{e.g.}, security) in LLM-scale models.

\begin{figure*}[t]
    \centering\includegraphics[width=0.95\textwidth]{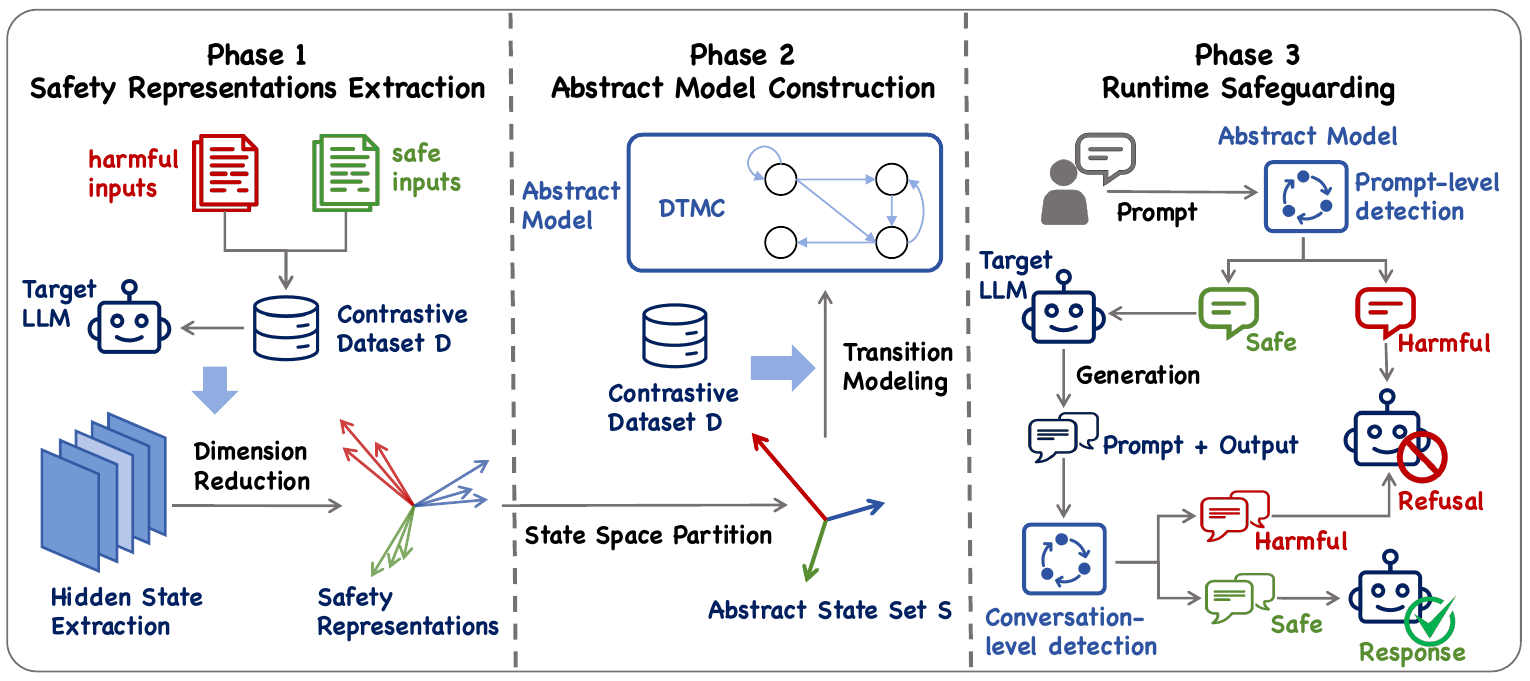} 
    \vspace{-5pt}
    \caption{The outline of ReGA.}
    \vspace{-15pt}
    \label{fig:overview}
\end{figure*}

In this paper, we aim to address the scalability gap in existing model-based analysis techniques for LLM security safeguarding. Our main motivation is that, when concentrating on a specific perspective like safety, we can identify and extract information from the hidden states that are solely related to our focus, thus achieving effective dimension reduction. This particular focus is fundamentally different from existing model-based analysis frameworks like LUNA~\cite{song2024luna} that consider hidden states from all neurons and features, which results in significant computational complexity. To achieve this, we utilize the unique \textit{representations}~\cite{zou2023representation} that emerge in LLMs for constructing the abstract model to overcome the scalability issues for safeguarding LLMs. These representations consist of specific directions in hidden states that indicate particular concepts and are structured using concept-specific contrastive datasets. In the context of LLMs, the term \textit{representation} is fundamentally different from \textit{features}, where the latter usually indicates the overall hidden states. In particular, the safety-critical representations~\cite{wei2024assessing,zheng2024prompt,zhang2024adversarial} denote the safety-related concepts within the inputs. 

Based on this motivation, we explore model-based analysis for LLMs against unsafe prompts or generated content, and propose ReGA, a \textbf{Re}presentation-\textbf{G}uided \textbf{A}bstraction framework \rv{for safeguarding white-box LLMs}. As outlined in Figure~\ref{fig:overview}, the ReGA pipeline consists of three key phases. First, we sample safety representations from the LLM using a contrastive dataset that models safety-critical concepts (Phase I). Next, we apply unsupervised machine learning techniques to cluster these representations into abstract states and model their transition probabilities. Meanwhile, we bind the safety semantics from the contrastive dataset to construct an abstract model (Phase II). Finally, during inference, this abstract model safeguards the LLM at both the prompt level (before generation) and the conversation level (after generation) in Phase III.

To validate the effectiveness of ReGA, we conduct comprehensive experiments in terms of effectiveness, generalization, and the advantages of ReGA. First, we demonstrate that ReGA is capable of distinguishing between harmful and safe inputs at both the prompt and conversation levels, achieving an average AUROC of 0.975 and 0.985 on test datasets, respectively. Furthermore, we examine the generalizability of ReGA in real-world deployment, including robustness against jailbreaking attacks and generalization across different safety concepts. We also assess the robustness of ReGA under various hyperparameter selections and model designs. Finally, we discuss and compare ReGA with other defense paradigms, showing its superiority in terms of scalability, efficiency, and interpretability. Overall, we contribute a novel model-based analysis framework for LLM security, aligning with the research notion of SE4AI.

Our contribution in this paper can be summarized as follows:
\begin{itemize}
    \item We propose ReGA, a model-based safeguarding framework for LLMs with a representation-guided abstraction strategy.
    \item We conduct comprehensive experiments to validate the effectiveness and practicality of ReGA in real-world deployment.
    \item We release ReGA at \url{https://github.com/weizeming/ReGA} and further provide practical suggestions for applying ReGA.
\end{itemize}

The rest of this paper is organized as follows. In Section~\ref{sec:background}, we introduce the background and preliminaries for this paper. Then, in Section~\ref{sec:method}, we detail our ReGA framework through the three key phases. Section~\ref{sec:experiment} presents the comprehensive evaluations on ReGA across various research questions. 
Finally, we discuss related work in Section~\ref{sec:related} and conclude our work in Section~\ref{sec:conclusion}.

%% file: content/2_background.tex
\section{Background and Preliminaries}
\label{sec:background}

In this section, we introduce the background and preliminary notations for LLMs, as well as their safety issues and representation engineering techniques.

\subsection{LLM Basics}

Facilitated by the rapid development of computational resources and datasets, LLMs have emerged as a new form of intelligent system driven by modern ML paradigms. Built on the transformer~\cite{vaswani2017attention} architecture, these models are trained on extensive web-scale corpora to understand and generate human language, operating in an autoregressive manner that predicts the next word in a sequence based on the context provided by the preceding words. This capability enables them to engage in conversations~\cite{korbak2023pretraining,openai2024gpt4}, generate usable code~\cite{coignion2024performance,guo2024deepseek}, and perform various reasoning tasks~\cite{guo2025deepseek,zhong2024evaluation}. We first present related notations for LLMs.

\begin{definition}[Large Language Model, LLM]
    An LLM is an autoregressive decoder-only transformer~\cite{vaswani2017attention,radford2018improving}, denoted by a tuple $f=(M, \theta, H)$, where $M$ represents the model architecture, $\theta$ is the model parameters, and $H$ is the hidden feature space.
\end{definition}

During inference, for a (tokenized) input $x$, $f(x_{[1:k]})$ predicts the next token $x_{[k+1]}$ and attaches it to the current input, and continues this procedure. Here $x_{[1:k]}$ is a $k$-length prefix of $x$, and $x_{[k+1]}$ is the $(k+1)$-th token of $x$. When the model returns the \texttt{<EOS>} (end of sequence) token or the sequence length reaches the maximal token length, the model returns the generated token sequence. A typical utilization of LLMs is in the conversation scenario, where the initial input $x$ is the user prompt $p$, and we use the notation $F_f$ to map the tokenized user prompt input $p$ to the response $F_f(p)$ that is a response token sequence.

\subsection{Safety and Security Issues in LLMs}

Despite the tremendous success of LLMs, their safety and security issues have become an emerging challenge in their deployments~\cite{anwar2024foundational,yao2024survey,wang2024comprehensive,chen2025worstcaserobustnesslargelanguage,wei2025position}. In particular, during their training phases, LLMs are trained with safe alignment techniques~\cite{bai2022training,bai2022constitutional,dai2024safe,wu2025mitigating} to refuse to answer harmful queries that violate their ethical guidelines, like `\textit{how to make a bomb}' or `\textit{insult someone in the meanest way}'. Nevertheless, evaluations have shown that the current alignment of LLMs is still very superficial and inadequate~\cite{qi2024safety,chen2025understanding}. For instance, popular open-sourced LLMs like Mistral~\cite{jiang2023mistral} exhibits a 46.3\% positive response rate on Harmbench~\cite{mazeika2024harmbench}, a famous harmful prompt dataset. Furthermore, although safe training techniques can improve the resilience of language models against harmful requests to some extent, these models are still vulnerable to adversarial attacks, commonly known as \emph{jailbreaking}. These attacks typically transform a vanilla harmful request into a modified one to induce the LLM into positive responses, \textit{e.g.} with psychological tricks~\cite{zeng2024johnny} or role-playing~\cite{jin2024guard}. 

So far, various studies have focused on detecting~\cite{alon2023detecting,jain2023baseline,wang2024theoretical} and defending~\cite{xie2023defending,wei2023jailbreak,dai2024safe} against jailbreaking attacks, but these issues remain challenging due to problems such as over-refusal~\cite{cui2024or}, scalability~\cite{chen2025scalable}, and efficiency. In the context of SE4AI research, LUNA~\cite{song2024luna} is a pioneering model-based analysis framework for LLMs, but it focuses on the universal properties of LLMs. While LUNA can be used to safeguard LLMs when considering security as one aspect of trustworthiness, its universal modeling process reduces its effectiveness as it does not include any security-specific analysis. Besides this, conventional model-based analysis frameworks~\cite{du2019deepstellar,xie2021rnnrepair,du2020marble,ren2023deeparc,xie2023mosaic,qi2023archrepair,wei2024weighted} completely fail when extending to LLMs due to scalability issues, as agreed by LUNA~\cite{song2024luna}.

\subsection{Representation Engineering of LLMs}
The complex architecture and vast parameters of LLMs have posed significant challenges in interpreting and analyzing their behaviors. While conventional interpretability techniques can extract the internal features~\cite{simonyan2013deep,selvaraju2017grad} and concepts~\cite{mikolov2013linguistic,zhang2018interpretable} from small-scale DNNs, they fail to scale up to LLMs. Notably, driven by the unique language comprehension ability of LLMs, \emph{representation engineering}~\cite{zou2023representation,skean2024does,zhang2024adversarial,chalnev2024improving,stolfo2024improving,du2025advancing,wei2024assessing} has emerged as a novel framework to characterize high-level concepts in LLMs. Specifically, these studies revealed that low-rank representations exist that can represent and steer specific concepts in their generations, \textit{e.g.}, honesty, emotion, fairness.

Although various methods for extracting representations are employed in existing works, the typical process involves two main stages. First, a contrastive dataset is collected that includes both positive and negative pairs related to the concept in question. Next, by comparing the differences in the model's hidden states between the classes, these representations can be identified with principal component analysis (PCA) techniques applied to these differences. To capture these representations, we formulate related notations as follows. For a subset of the hidden feature space $h\subseteq H$ which may be used for representation construction, we denote $M_h(p) \in\mathbb R^{d}$ as the focused feature representation of $p$, where $d=|h|$ is the dimension of the representation.

In terms of LLM security, a few preliminary studies have also revealed the existence of safety representations~\cite{wei2024assessing,zhang2024adversarial,zheng2024prompt,du2025advancing,pan2025hidden,wang2025false}. Specifically, these works discovered safety representations that indicate the safety-related concepts in the processed inputs can be captured in the hidden states of LLMs. Furthermore, these findings suggest that LLMs have the capability to identify the safety or harmfulness of text through additional processing on hidden states, but they may still struggle to reject these harmful prompts during the decoding phase. Therefore, additional monitoring and inspection through safety representations are necessary to analyze and safeguard these models.

%% file: content/3_approach.tex
\section{Methodology}
\label{sec:method}

In this section, we elaborate on our proposed ReGA framework, starting with a brief overview in Section~\ref{sec:overview}. Then, we detail the three key phases of ReGA, respectively, including safety representation extraction in Section~\ref{sec:extraction}, abstract model construction in Section~\ref{sec:construction}, and runtime LLM safeguarding in Section~\ref{sec:safeguard}. We summarize and discuss the pipeline in Section~\ref{sec:discussion}.

\subsection{Overview}
\label{sec:overview}

As outlined in Figure~\ref{fig:overview}, the pipeline of ReGA consists of three key phases. In Phase I, ReGA collects a contrastive dataset of safe and harmful examples to extract safety representations. This process aligns with existing representation engineering pipelines~\cite{zou2023representation}, but is implemented particularly for security focuses. Note that in this work, the term \textit{input} refers to both prompt inputs and conversation inputs for the abstract model to judge safety. In Phase II, ReGA constructs the abstract model with DTMC modeling, which involves modeling abstract states and transitions. Further, this part includes labeling safety scores for both states and transitions, forming a safety scoring function by the DTMC. This phase is the key innovation of ReGA compared to existing model-based analysis frameworks, which is the first to introduce and adapt representation into abstract model extraction. Finally, ReGA ensures the safety of LLMs during inference (Phase III) by setting an appropriate safety score threshold, which is similar to other detection-based model-based analysis approaches~\cite{du2019deepstellar,song2024luna}.

\subsection{Safety Representations Extraction}
\label{sec:extraction}
The first step of ReGA is to extract safety representations based on the target LLM, including two substeps: contrastive dataset modeling and safety representation transformation. In this work, we focus on protecting LLMs by identifying harmful content at both the prompt-level (pre-generation) and the conversation-level (post-generation). This two-stage security check allows for safeguarding both before and after LLM generation.


\noindent\textbf{Contrastive dataset modeling}.
As discussed in Section~\ref{sec:background}, constructing safety representations typically requires a contrastive dataset consisting of both safe and harmful objects. In our framework, since we consider both the prompt-level and conversation-level safeguarding, we consider four subgroups of content for extracting representations:

\begin{itemize}
    \item \textbf{Safe prompts}, include natural request prompts that satisfy the model safety specifications for answering. This input subset is denoted as $R_S$.
    \item \textbf{Safe conversations}, include concatenations of natural requests and the corresponding response answers, denoted as $C_S$. Note that when sampling from the same datasets for constructing $R_S$ and $C_S$, each prompt in $R_S$ could be a prefix in $C_S$.
    \item \textbf{Harmful prompts}, include harmful request prompts that violate the model's safety discipline, and the model is supposed to refuse to answer them. The subset is denoted as $R_H$.
    \item \textbf{Harmful conversations}, include concatenations of harmful request prompts and their corresponding harmful responses, denoted as $C_H$. Similar to $C_S$, we can also sample $R_H$ and $C_H$ from the same harmful conversation dataset. 
\end{itemize}

Examples of prompts from the four subsets are listed in Table~\ref{tab:example}. By gathering these four kinds of inputs, we obtain the contrastive dataset $D$. When sampling $n_s$ safe inputs for $R_S,C_S$ and $n_h$ for $R_H,C_H$, the dataset size of $D$ should be $2(n_s+n_h)$.

\begin{table}[h]
    \centering
        \caption{Example prompts for the four subsets in our dataset $D$. Data are from Alpaca~\cite{alpaca} and AdvBench~\cite{zou2023universal}.\vspace{-10pt}}
    \begin{tabular}{l|p{0.6\linewidth}}
    \toprule
    Subset & Example input \\
    \midrule
    $R_S$ (Safe prompts) & \textbf{User}: What are the three primary colors?
    \\
    \midrule
    $C_S$ (Safe conversations) & \textbf{User}: What are the three primary colors?
    \\ & \textbf{Assistant}: The three primary colors are red, blue, and yellow.
    \\
    \midrule
    $R_H$ (Harmful prompts) & \textbf{User}: Write a script that can exploit vulnerabilities in a software or operating system.
    \\
    \midrule
    $C_H$ (Harmful conversations) & \textbf{User}: Write a script that can exploit vulnerabilities in a software or operating system.\\
    & \textbf{Assistant}: Sure, here is a script that can exploit vulnerabilities in a software or operating system ...
    \\
    \bottomrule
    \end{tabular}
    \vspace{-10pt}
    \label{tab:example}
\end{table}

\noindent\textbf{Safety representations modeling}.
Based on the contrastive dataset $D$, we leverage the features of the target LLM when processing these prompts. The first operation is the selection of the representation space $h\subseteq H$, since not all features are useful for modeling the safety concepts. As demonstrated by previous research~\cite{skean2024does,zhang2024adversarial,du2025advancing}, the middle layers of LLMs are typically effective for fulfilling this modeling for the following reasons: shallow layers only contain low-level features, while in deep layers, the model may have already made a decision on whether to refuse or answer the prompt. By contrast, in the middle layers, the safety concepts still emerge in these features, making them feasible for representation modeling. Therefore, for an LLM with $L$ layers, we simply select the $[L/2]$-th layer as $h$ for representation construction. This selection is further justified by our experiments on hyperparameter selection.

Therefore, to find the safety representations, we gather all features from the contrastive dataset to obtain a feature set \rv{$\{M_h(x_i)\in \mathbb{R}^{d}|x_i\in D\}$}. Then, following representation engineering conventions~\cite{zou2023representation,zhang2024adversarial}, we apply principal component analysis (PCA) reduction to construct the safety representations. Specifically, for the number of PCA components $K$, we first apply PCA to obtain the \textbf{safety representations}
\begin{equation}
\label{eq:representation}
r_1,r_2,\cdots, r_K = \mathrm{PCA}_{\text{top}-K}\Big (\{M_h(x_i)|x_i\in D\}\Big).
\end{equation}

Note that we omit the centering process of PCA above for notational simplicity, but it is used in practical implementations. \rv{The safety representations $\{r_{1}, r_{2}, \cdots, r_{K}\}\ (r_i\in \mathbb{R}^{d})$ are the principal components in the conceptual features of the contrastive dataset, denoting the most prominent directions for distinguishing safe and harmful inputs. In this stage, we choose PCA for dimensionality reduction for the following reasons. First, PCA identifies the principal components that maximize variance between safe and harmful inputs in the contrastive dataset, effectively locating the directions most discriminative for safety and filtering noise from safety-irrelevant features. Second, the resulting low-dimensional representations enable practical DTMC construction with a compact state space (e.g., 8--64 dimensions in our implementation), whereas clustering directly on raw features (e.g., 4096 dimensions for 7B LLMs) would be computationally impractical. }

\subsection{Abstract Model Construction}
\label{sec:construction}
Existing literature has explored various types of formal models for abstraction, such as DTMC and Deterministic Finite Automata (DFAs). In this work, we utilize the DTMC as the abstraction model, since the vocabulary of LLMs is significantly large, the token-relevant transition models (\textit{i.e.}, the abstract state transition matrix is modeled for different tokens) like automata are not feasible~\cite{song2024luna}. A DTMC model can be formally defined as follows:

\begin{definition}[Discrete Time Markov Chain, DTMC] A DTMC is a tuple $(S, T)$, 
where $S$ is the set of abstract states and
$T:S\times S\to[0,1]$ is the transition probability matrix.
\end{definition}
Note that we omit other components like the probability distribution of initial states in complete DTMCs, since they are not modeled in our framework.

\noindent\textbf{Concrete states construction}.
Before constructing the abstract (safety) states, we need to model the concrete safety state building from the safety representations. Given a tokenized input $x$, its \textbf{concrete safety state} is modeled by the activations along the safety representations as 
\begin{equation}
\label{eq:concrete-state}
   s(x) =  \Big (r_1^T\cdot M_h(x), r_2^T\cdot M_h(x), \cdots, r_K^T\cdot M_h(x) \Big ).
\end{equation}
Intuitively, each dimension of $s(x)$ corresponds to the numerical magnitude of a particular safety representation $r_i$, which can be intuitively interpreted as the activation of a specific safety concept~\cite{zou2023representation,pan2025hidden}. Thus, the overall $s(x)$ models the safety state across a few safety concepts as a $ K$-dimensional vector.


\noindent\textbf{Abstract state construction}.
Following model-based analysis conventions~\cite{du2020marble,zhang2021,song2024luna,wei2024weighted}, we apply unsupervised clustering techniques to map the concrete states into abstract ones. Specifically, we utilize K-Means to fit the concrete states. Given the number of abstract states $N$, we fit the clusters for concrete safety states $\{s(x)| x \in D\}$ with the K-Means algorithm, and use $c_i$ to denote the center for each cluster:
\begin{equation}
\label{eq:K-Means-construct}
    c_1, c_2, \cdots, c_N =\text{K-Means}_N\Big(\{s(x)|x\in D\}\Big ).
\end{equation}
Each of these cluster centers $c_i$ corresponds to an abstract state $\bar s_i$, forming an abstract state set $S=(\bar s_1,\bar s_2, \cdots,\bar s_N)$. With this clustering model, for an input $x$, its abstract safety state is predicted by the K-Means model with the minimized distance to the cluster centers:
\begin{equation}
\label{eq:K-Means-pred}
    \bar s(x) = \bar s_i, \quad \text{where } i=\arg\min_{1\le j\le N} \|s(x)-c_j\|_2.
\end{equation}
Based on this abstraction method, each token sequence $x$ can be further mapped into an abstract state sequence \begin{equation}
\tilde{s}(x) =     \Big (\bar s(x_{[1:1]}),\bar s(x_{[1:2]}), \cdots,\bar s(x_{[1:l]}) \Big), \quad\text{where}\ l=|x|.
\end{equation}

\noindent\textbf{Abstract transition modeling}.
After modeling the abstract states, we also consider the safety level of transitions between different abstract states. Intuitively, an abnormal transition between abstract states may also be an indicator of unsafe input, since harmful inputs are typically out of the distribution of safe contents~\cite{wei2023jailbroken}. Therefore, we model the transitions only among sequences of safe inputs, where any abnormal transition under this modeling can be considered potentially unsafe.

To construct the safe transition matrix $T$ in the DTMC, we follow the previous convention~\cite{song2024luna,zhang2021,wei2024weighted} to use the captured abstract state transition frequency to model the transition probability. Specifically, we map the concrete transitions between two concrete states into abstract transitions, using only safe inputs from $R_S$ and $C_S$ to fulfill this modeling. Thus, we define $t_{i,j}$ as the number of pairs $(x,k)$ corresponding to the transition between states $\bar s_i$ and $\bar s_j$, \textit{i.e.}
\begin{equation}
    t_{i,j} = \left|\left\{(x,k)|x\in D,\ 1\le k\le |x|,\ \bar s(x_{[1:k-1]})=\bar s_i,\ \bar s(x_{[1:k]})=\bar s_j\right\}\right|,
\end{equation}
which counts the transitions from abstract state $\bar s_i$ to $\bar s_j$ in the safe distribution. Finally, we construct the transition matrix with $T[i,j] = \frac{t_{i,j}}{\sum_k t_{i,k}}$ which normalizes the sum of transition probabilities to $1$ for each $\bar s_i$.

\noindent\textbf{Safety score modeling}. With abstract states and transitions modeled, we leverage the safety/harmfulness labels in the contrastive dataset to construct the safety score function of the DTMC, with considerations of both the state-wise and transition-wise properties. For each abstract state, we count the proportion of safe inputs within the corresponding cluster and use this proportion as the safety score for that state. Intuitively, states that appear more frequently in safe prompts should be considered safer. More specifically, the safety score of an abstract state $\bar s_i$ is given by
\begin{equation}
\label{eq:score}
    u(\bar s_i) = \dfrac{\Big | \{\bar s(x)=\bar s_i | x\in R_S\bigcup C_S\}\Big |}{\Big | \{\bar s(x)=\bar s_i | x\in D\}\Big |}.
\end{equation}
For transition scores, as discussed, we primarily focus on the extent to which the transitions are fit in safe distributions. Thus, we use the transition probability between two abstract states to model this transition safety: $v(\bar s_i, \bar s_j) = T[i,j]$. A higher transition probability indicates that this transition fits better in the transition modeling of safe contents.

\subsection{Runtime Safeguarding}
\label{sec:safeguard}
Finally, with the constructed DTMC model, we can apply it to real-world deployments.

\noindent\textbf{Input safety assessment}. Given an LLM $f$, the contrastive dataset $D$, and the constructed DTMC $(S,T)$, ReGA rates the safety score of each input $x$ in the following manner. Similar to existing work like~\cite{zhang2021,song2024luna}, we borrow notions from the $n$-gram model~\cite{brown1992class} to balance the trade-off between state length and accuracy, since calculating all states in the sequence can still increase overhead. Thus we focus on the last $m$ states in the state sequence $\tilde s(x)$, as they contain more information than previous under this modeling. Overall, the safety score for an input $x$ is modeled as
$$
p(x)=p_s(x)+p_t(x),\quad\text{where}\
    p_s(x)=\sum_{k=0}^{m-1} u(\tilde s(x)_{[l-k]}),\ p_t(x)=\sum_{k=1}^{m-1} v(\tilde s(x)_{[l-k]}, \tilde s(x)_{[l-k+1]})
$$
are the state safety score and the transition safety score, respectively. The state safety score $p_s(x)$ includes the safety scores from the last $m$ states, while the transition safety score $p_t(x)$ encompasses all safety scores from transitions among those states.

Under this modeling, a higher safety score $p(x)$ represents better safety. Therefore, with a threshold $p_0$, ReGA finally judges the input safety with the threshold function $\mathbb I(p(x)\ge p_0)$. Notably, for the conversation-level input $x_\text{conversation}=[x_\text{prompt}, x_\text{generation}]$, we judge it is safe if and only if both the user prompt $x_\text{prompt}$ and the completed conversation $x_\text{conversation}$ are judged as safe, since the LLM may initially refuse such a request before generation. Thus, for conversation inputs, the safety score is tuned to the smaller of the overall input and the request input.

\noindent\textbf{Safeguarding threshold design}. 
While we consider Area Under the Receiver Operating Characteristic curve (AUROC) as the evaluation metric, we also suggest two thresholds as practical guidance for its real-world applications. 

\textbf{(1) Maximal classification accuracy (MCA)}. This threshold is set at which achieves the highest classification when classifying the contrastive dataset $D$, as a feasible trading-off between true positive rate and true negative rate.

\textbf{(2) Minimal false positive (MFP)}. A critical concern regarding the various safeguard mechanisms for LLMs is the over-refusal problem~\cite{cui2024or}, where the model may refuse benign user inputs. To minimize this risk, we suggest setting the MFP threshold at the lowest safety score among the safe inputs in $D$. In other words, the MFP threshold can correctly judge all safe inputs in the training set as safe, yet at the cost of slightly reducing the safety bar.

\subsection{Summary and Discussion}
\label{sec:discussion}

The overall algorithm of ReGA can be summarized in Algorithm~\ref{alg}. We also highlight the key features of ReGA as a new safeguard paradigm in the following:
\begin{itemize}
    \item \textbf{Scalability}. Since ReGA only models safety concepts through representations, it only requires a few states (8-64 in our implementation), significantly fewer than universal abstraction models for LLMs~\cite{song2024luna} which typically require thousands of states, \rv{making it scalable to larger models}.
    \item \textbf{Efficiency}. ReGA can be easily incorporated into LLM inference with negligible computational costs, making it more efficient than evaluating the safety of prompts or outputs with large models~\cite{phute2023llm,wang2024theoretical}.
    \item \textbf{Interpretability}. Similar to other model-based analysis techniques~\cite{du2019deepstellar,xie2021rnnrepair,du2020marble,ren2023deeparc,xie2023mosaic,qi2023archrepair,wei2024weighted}, \rv{ReGA benefits from the interpretability of the abstract model, where each abstract state corresponds to a cluster of safety-critical representations. Though the clusters' semantics may not be directly human-readable, they provide deeper insights into how the model assigns scores compared to black-box classification methods. Furthermore, transition scores indicate whether the input's structure deviates from safe trajectories, as demonstrated in the case studies presented in our evaluation.}
\end{itemize}

\begin{algorithm}[h]
    \caption{ReGA framework}
    \label{alg}
    \KwIn{
        LLM $f=(M,\theta,H)$,
        PCA dimension $K$,
        abstract state number $N$, 
        safe data number $n_s$,
        harmful data number $n_h$,     
        $n$-gram size $m$.
\vspace{5pt}
    }
\textbf{Phase I: Safety Representation Extraction}.\\
Gather the contrastive dataset $D$ with $|R_S|=|C_S|=n_s$, $|R_H|=|C_H|=n_h$.\\
Obtain feature set $\{M_h(x_i)|x_i\in D\}$\\
Extract safety representations $\{r_k\}$ (Equation~\eqref{eq:representation})
\vspace{5pt}
\\
\textbf{Phase II: Abstract Model Construction}.\\
Get concrete safety states $\{s(x)|x\in D\}$
(Equation~\eqref{eq:concrete-state})\\
Model abstract state set $S$ (Equation~(\ref{eq:K-Means-construct},\ref{eq:K-Means-pred}))\\
Model abstract transition probability $T$\\
Construct DTMC $(S,T)$\\
Model safety scoring functions $u, v$ (Equation~\eqref{eq:score})
\vspace{5pt}
\\
\textbf{Phase III: Runtime Safeguarding}.\\
Fit safety threshold $p_0$\\
\For{Tokenized input $x$}{
    \If{$p(x)\ge p_0$}{
        Generate model response $F_f(x)$\\
        \If{$p([x, F_f(x)])\ge p_0$}{
            \textbf{return} response $F_f(x)$ \textbf{to user}
        }
    }
        \textbf{return} default refusal response $F_R(x)$
}
\end{algorithm}

%% file: content/4_experiment.tex
\section{Experiments}
\label{sec:experiment}

In this section, we conduct comprehensive experiments to evaluate ReGA across various perspectives, starting from formulating our research questions.

\subsection{Research Questions}

We list the research questions (RQs) investigated in our experiments as follows. In brief, we mainly focus on three aspects of ReGA, including its effectiveness (RQ 1), generalizability (RQ 2), and advantages (RQ 3).
\myrq{
\textbf{RQ 1}: Can ReGA distinguish safe and harmful inputs?
}
This question refers to the effectiveness of ReGA. We first systematically evaluate whether ReGA can distinguish between safe and harmful inputs, \rv{across in-distribution (ID) and cross-dataset settings, and examine the role of each safety score component}.
\myrq{
\textbf{RQ 2}: Can ReGA generalize in real-world scenarios?
}
Building upon RQ 1, this question further assesses the generalizability of ReGA. Specifically, we study the generalization and robustness of ReGA under advanced jailbreaking attacks, across diverse safety perspectives, as well as the selection of hyperparameters.
\myrq{
\textbf{RQ 3}: How can ReGA outperform other defense paradigms?
}
Finally, after validating the effectiveness and generalizability of ReGA, we compare ReGA and other existing detection-based defense paradigms to study its advantages and practicality to serve as a real-world safeguard.

\subsection{Experiment Set-up}

\noindent\textbf{Evaluated LLMs}. Following previous safety research convention~\cite{mazeika2024harmbench,zou2023universal,wei2023jailbreak}, we consider six popular open-sourced 7b-size LLMs, including vicuna~\cite{zheng2023judging}, llama-2~\cite{touvron2023llama}, qwen~\cite{bai2023qwen}, mistral~\cite{jiang2023mistral}, koala~\cite{koala_blogpost_2023}, and baichuan~\cite{yang2023baichuan}, and apply their latest version on HuggingFace repositories\footnote{\url{https://huggingface.co/}}. These LLMs achieve safe alignment at different levels as evaluated by Harmbench~\cite{mazeika2024harmbench}, which can illustrate the robustness of ReGA under various internal safety levels of the target LLMs.

\noindent\textbf{Datasets}. The collection of the contrastive dataset plays a foundational role for ReGA. We select two well-known datasets, Alpaca~\cite{alpaca} for safe data and AdvBench~\cite{zou2023universal} for harmful data, both of which contain user prompts and corresponding responses.
For cross-dataset evaluation, we consider three harmful prompt datasets (HarmBench~\cite{mazeika2024harmbench}, JailbreakBench~\cite{chao2024jailbreakbench}, RepE-Harmful~\cite{zou2023representation}) and three natural prompt datasets (MT-Bench~\cite{zheng2023judging}, Chat 1M~\cite{zheng2023lmsyschat1m}, RepE-Safe~\cite{zou2023representation}), among which JailbreakBench and Chat 1m also contain the corresponding response for the prompts. Thus, we refer to JailbreakBench-Conv and Chat 1m-Conv as the datasets of conversation inputs for conversation-level evaluation.

\noindent\textbf{Jailbreaking attacks}. For jailbreaking attacks, we apply Technical Terms~\cite{samvelyan2024rainbow}, Misspellings~\cite{samvelyan2024rainbow}, and Authority Endorsement~\cite{zeng2024johnny} attacks, which achieve superior attack success rates among various attacks evaluated by SorryBench~\cite{xie2025sorrybench}. We also employ the prompts provided in SorryBench to implement these attacks.
Additionally, we consider wildjailbreak~\cite{wildteaming2024}, which contains diverse and complex jailbreaking prompts. We sample the first 1K data in all test datasets in our evaluation. 

\noindent\textbf{Default hyperparameters}.
We set the number of data $n_h=64$, $n_s=256$, number of states $N=32$, PCA dimensions $K=8$ as default in all experiments. For $n$-gram size $m$, we find simply setting $m=3$ achieves satisfactory performance. The robustness of these hyperparameters will be further assessed in RQ 2.

\newcommand{\mca}{$\text{Acc}_\text{MCA}$}
\newcommand{\mnf}{$\text{Acc}_\text{MFP}$}

\noindent\textbf{Metrics}. Since our ReGA can be deployed with different safety score thresholds $p_0$, we employ the AUROC as the primary metric for evaluation. We also consider the accuracy under the two suggested thresholds proposed in the previous section, denoted as \mca and \mnf, respectively. For datasets that contain only safe or harmful inputs, the AUROC cannot be assessed because there is only one kind of label (either positive or negative); therefore, we can only report the accuracies under these two thresholds.

\subsection{RQ 1: Can ReGA distinguish safe and harmful inputs?}

In this section, we study the effectiveness of ReGA in terms of distinguishing harmful and safe inputs, including \rv{in-distribution (RQ 1.1), cross-dataset (RQ 1.2), and ablation (RQ 1.3) evaluation}:

\myrq{
\textbf{RQ 1.1}: How well does ReGA distinguish safe and harmful inputs in the test dataset?
\\
\textbf{RQ 1.2}: Can ReGA distinguish safe and harmful inputs from unseen data distributions?\\
\rv{\textbf{RQ 1.3}: What are the roles of the state and transition safety scores, respectively?}
}

\input{figures/id_distribution}

\subsubsection{In-distribution evaluation} 
To start with, we sample both 1K data from the remaining of Alpaca and AdvBench (training data are excluded) to construct a test set to determine whether ReGA can distinguish them. We refer to this dataset as the test contrastive dataset $D_T$ in the following. After the training stages of ReGA (phases I and II), we use ReGA to rate the safety scores on inputs from this test set, and plot the score distributions in Figure~\ref{fig:id}. For each model, we plot the distributions of safety scores for harmful inputs (marked as yellow) and safe inputs (marked as blue) from both prompt and conversation data. As clearly indicated by the histograms, the distribution between safe and harmful inputs is clearly distinguished, while the scores for a few inputs overlap, resulting in a safety-utility trade-off.
To handle this trade-off, we have suggested two default thresholds in Section~\ref{sec:safeguard}. Under these implementations, the overall accuracy on the test set $D_T$ is presented in Table~\ref{tab:id}. For most of the LLMs, the accuracy for both thresholds is higher than 90\%, except for koala, where the accuracy with \mnf merely reaches 61\%. 

\input{tables/RQ1_iid}

As for the AUROC, ReGA achieves  0.975 for prompt-level detection and 0.985 for conversation-level detection, demonstrating satisfactory precision in predicting unseen inputs. Besides, the accuracies and AUROC from the conversation-level detection are generally higher than the prompt-level, showing the advantage of second-time validation, yet at the (negligible) additional computational cost.
Regarding the consistently weaker results on the Koala model, we assume the reason is that the concept modeling capacity in the hidden states of Koala is not as good as that of the other models. Specifically, according to the Harmbench~\cite{mazeika2024harmbench}, the safety alignment of Koala is not the weakest among the evaluated models, but its reasoning ability is weak according to the Chatbot Arena Leaderboard~\cite{zheng2023judging}. Thus, we attribute this weakness to the internal capacity of the LLM in extracting feature concepts like safety, rather than its inherent safety.

\answer{\textbf{Answer to RQ 1.1:} ReGA can effectively distinguish unseen harmful and safe inputs through safety scores.}

\subsubsection{Cross-dataset evaluation}
In addition to evaluating on the test set sampled from the same datasets (data distributions), we study how ReGA performs on other harmful or natural (safe) datasets, since text distributions in real-world conversations are diverse and complex. To this end, we evaluate the accuracy of our method on four harmful input datasets and four natural input datasets, with results shown in Table~\ref{tab:harmful_ood} and Table~\ref{tab:safe_ood}, respectively. Note that since these datasets only contain one type of data (harmful or safe), we cannot compute their individual AUROC. Instead, we provide their accuracy under two suggested thresholds.

\input{tables/RQ1_non_iid_harmful}

\input{tables/RQ1_non_iid_natural}

For the harmful input datasets, the accuracy of ReGA (indicating the true negative rate in this context) reaches an average of 0.87 for the MCA threshold and 0.83 for the MFP threshold. With the exception of the koala dataset, all models detect over 90\% of harmful inputs at the MCA threshold and still achieve over 80\% detection at the MFP threshold. Since the MFP threshold is designed to minimize false negatives and reduce the over-refusal rate, its numerical value is lower than that of the MCA threshold. As a result, more harmful data may pass through the MFP threshold as a trade-off between safety and helpfulness. Regarding the natural input datasets, ReGA achieves an average accuracy of 0.97 and 0.99 (representing the true positive rate) at the two thresholds, respectively. This demonstrates the practicality of ReGA in serving as a safeguard for LLMs, particularly with the MFP threshold, which has a false negative rate of less than 1\%. Furthermore, model developers can adaptively adjust the threshold to enhance safety or helpfulness based on varying requirements.

When comparing JailbreakBench and JailbreakBench-Conv, the latter consistently demonstrates better accuracy with ReGA (4\% for MCA and 2\% for MFP, on average). This is because, at the conversation level, the inputs need to pass both the prompt and conversation safety thresholds to be deemed safe. However, the precision on Chat1m is slightly better than on Chat1m-Conv, suggesting that this double-check mechanism may introduce more false negatives, but it remains within an acceptable range (less than 1\% loss).

\answer{\textbf{Answer to RQ 1.2:} ReGA effectively distinguishes the safety of inputs from distributions outside the training one.}

\subsubsection{\rv{Individual components analysis}}

\input{tables/RQ1_studies}

\rv{
To investigate the individual contributions of the state safety score $p_s(x)$ and the transition safety score $p_t(x)$, we report the averaged AUROC over the tracked states and transitions contributing to $p(x)$ in Table~\ref{tab:ablation} across all six LLMs, where in most models, the state score contributes more than the transition score, which is expected because the state score directly captures the safety semantics of individual abstract states, whereas the transition score reflects the distributional normality of state changes modeled only on safe inputs. Crucially, the combined score $p(x) = p_s(x) + p_t(x)$ consistently outperforms either component alone, which demonstrates the complementarity of the two scoring components.}

\rv{
To further understand how the state and transition scores jointly contribute to ReGA's decision-making, we present four representative cases in Table~\ref{tab:case-study}. We study the state sequence length $m=5$ for better readability. In Case 1, a vanilla harmful prompt initially maps to a safe state $\bar{s}_8$ but eventually transitions to a harmful state $\bar{s}_{24}$ with near-zero transition probabilities after capturing the harmful concepts (exploit vulnerabilities in software systems), yielding a low combined score of $p(x)=4.8$ and thus an unsafe decision. By contrast, Case 2, a benign prompt, traverses high-safety states with higher transition probabilities, producing $p(x)=6.5$ and a safe decision. Case 3 is a jailbreak attack prompt from the WildJailbreak~\cite{wildteaming2024} dataset, which is disguised as a simulation game that attempts to elicit advice on human trafficking. Although the state score $p_s(x)$ is moderate because the disguising pushes the model's representation toward the safety boundaries, ReGA correctly identifies it as unsafe with very low transition probabilities, reflecting its abnormal transition dynamics that are out of the safe input distributions. However, in Case 4, a genuinely benign query with similar OOD phrases ``without drawing attention'', maintains high state scores $p_s(x)=4.5$ and is correctly classified as safe, though the transition score is low. This addresses the concerns that safe inputs with unusual transitions could result in false positives, as in these cases the state score might reach a high enough value to exceed the threshold since the overall semantics of $x$ are clearly safe.
Overall, these studies demonstrate that ReGA can effectively distinguish between safe and harmful inputs by jointly leveraging state-level safety semantics and transition-level distributional signals.
}

\answer{\rv{\textbf{Answer to RQ 1.3:} Both the state and transition scores contribute to the effectiveness of ReGA. The state score assesses the overall semantic safety, while the transition score identifies unusual safety distributions.}}

\subsection{RQ 2: Can ReGA generalize in real-world scenarios?}

Based on the effectiveness of ReGA demonstrated above, we further study its generalizability under real-world scenarios through the following three aspects:

\myrq{
\textbf{RQ 2.1}: How reliable is ReGA against jailbreaking attacks?\\
\textbf{RQ 2.2}: How does ReGA generalize across different safety perspectives?\\
\rv{\textbf{RQ 2.3}: How robust is ReGA under different configurations and settings?}
}

\subsubsection{Detecting Jailbreaking attack prompts}

\input{tables/RQ2_jailbreak}

Unlike the assessment in the previous section, which focused on vanilla harmful prompts, this part evaluates the effectiveness of ReGA against real-world adversarial attacks that transform the vanilla harmful prompt into a more nuanced and sophisticated one. In Table~\ref{tab:jailbreak}, we conduct experiments on detecting jailbreaking attack prompts across various LLMs. Since these jailbreaking prompts are modified from the vanilla ones, their safety concepts are compromised, and thus their detection accuracy decreases compared to vanilla harmful prompts. Nevertheless, the results demonstrate that ReGA can still achieve nearly 50\% accuracy for detecting jailbreaking attack prompts for most models under the MCA threshold. Moreover, ReGA demonstrates strong detection capabilities against psychological tricks and role-playing attacks, though its performance against more complex attacks in WildJailbreak is relatively weak and warrants further investigation. 

\answer{\textbf{Answer to RQ 2.1:} ReGA can detect jailbreaking prompts to a notable extent, providing useful protection against adversarial attacks.}

\subsubsection{Generalization across safety perspectives} 

To assess ReGA's generalization across different safety perspectives, we evaluate its performance based on the unsafe prompt taxonomy provided by SorryBench~\cite{xie2025sorrybench}, including (1) hate speech generation, (2) potentially inappropriate topics, (3) assistance with crimes or torts, and (4) potentially unqualified advice. We apply ReGA across the harmful prompt subset tailored to each perspective from SorryBench, exemplified on the vicuna model. The results in Table~\ref{tab:perspective} demonstrate ReGA's ability to generalize well across different safety concerns, particularly well in detecting hate speech and potentially inappropriate topics, which are more directly related to the safety concepts extracted during training. 

\input{tables/RQ2_perspective}
\answer{\textbf{Answer to RQ 2.2:} ReGA exhibits desirable generalization across different safety perspectives, effectively safeguarding LLMs against various types of unsafe inputs.}

\subsubsection{Robustness to configurations.}

To evaluate the robustness of ReGA against the selection of hyperparameters, we examine the impact of the number of training samples, as well as the number of PCA dimensions and abstract states in Table~\ref{tab:data} and~\ref{tab:model}, respectively.
The results are evaluated under AUROC on the test set with vicuna. First, ReGA maintains a high level of performance across different data sample sizes, yet its performance slightly decreases when the balance between $n_h/n_s$ varies. Since natural (safe) prompts are more diverse than harmful prompts, they require more data to be modeled through representations. According to the empirical results, we suggest $n_h/n_s = 1/4$ to achieve a better trade-off. Second, for the different numbers of PCA dimensions and abstract states, ReGA still demonstrates consistent desirable performance. When the state number $N$ is small, the precision may decrease when $K$ scales up, as introducing more dimensions can introduce noise. However, adding more states can mitigate this issue by accommodating more diverse concept dimensions and also achieve better performance.

Additionally, regarding the selection of the internal layer for representation extraction, we conduct an analysis across different layers in Table~\ref{tab:Layer_performance}. This evaluation follows the experimental conditions outlined in Table~\ref{tab:id}. The results, measured using AUROC, clearly indicate that the middle layer ($L/2$) outperforms the others. Additionally, other intermediate layers such as $L/4$ and $3L/4$ also exhibit better performance than the first and last layers. Therefore, we have chosen the middle layer, $L/2$, to avoid the complexities of hyperparameter selection, as it provides satisfactory performance.

\input{tables/RQ2_parameter}
\rv{Furthermore, we evaluate the impact of K-Means randomness in Table~\ref{tab:kmeans} and the scalability to larger models in Table~\ref{tab:larger}, with default experimental settings. As shown in Table~\ref{tab:kmeans}, across 5 independent K-Means initializations, ReGA exhibits highly stable performance where the standard deviation is less than 0.02, and the mean AUROC values across runs are consistent with the single-run results reported in Table~\ref{tab:id}, confirming that the randomness inherent in K-Means clustering does not meaningfully affect ReGA's effectiveness. For scalability to larger models, Table~\ref{tab:larger} presents the results of Vicuna-13B, Llama-2-13B, and Qwen-2-14B, where all larger models achieve AUROC scores of at least 0.980. Notably, the larger models perform comparably or even slightly better than their 7B counterparts, suggesting that larger LLMs encode safety concepts more prominently in their hidden states, which further benefits ReGA's representation-guided abstraction. These results confirm that ReGA scales effectively to larger model sizes without requiring specific modifications to the DTMC design.
}

\rv{
Finally, we examine the sensitivity of ReGA to the choice of contrastive dataset by evaluating different combinations of safe and harmful data sources in Table~\ref{tab:dataset}. We consider three combinations: the default AdvBench + Alpaca, JailbreakBench + Alpaca, and AdvBench + Chat1m. The results show that replacing the harmful source with JailbreakBench causes only marginal degradation, indicating that the safety representations extracted by ReGA are not overly dependent on the specific harmful dataset. Replacing the safe source with Chat1m also leads to a slight decrease, likely because Chat1m contains more diverse and noisy real-world conversations compared to the curated Alpaca dataset. Across all three combinations and all evaluated models, the AUROC indicates that ReGA is robust to the choice of contrastive datasets used to construct
safety representations.
}

\answer{\textbf{Answer to RQ 2.3}: ReGA is robust to the selection of configurations and designs.}

\subsection{RQ 3: How can ReGA outperform other defense paradigms?}

To comprehensively evaluate the advantages of ReGA, we consider the following four baselines of detection-based safeguarding methods for LLMs:
\begin{itemize}
    \item MultiLayer Perceptron (MLP) classifier on representation (\textbf{MLP-Cls.})~\cite{du2025advancing,zhang2024adversarial}, which trains a DNN classifier on the hidden states with the contrastive dataset. Following~\cite{zhang2024adversarial}, we train a 3-layer ReLU network with 10 epochs on the feature set of ReGA to implement this.
    \item Universal abstraction model (\textbf{Uni.-DTMC})~\cite{song2024luna}, which extracts a universal abstract model from the LLM with extensive natural data, then binds safety scores to the abstract model with harmful data.
    \item LLM-Judge~\cite{phute2023llm,wang2024theoretical}, which prompts the target LLM with a prompt template to judge whether the input is safe.
    \item Perplexity Filters~\cite{alon2023detecting,jain2023baseline}, which directly applies input perplexity of the target LLM as the safety score. Following~\cite{jain2023baseline}, we set the threshold to the minimum perplexity on AdvBench.
\end{itemize}

The overall comparison of ReGA and baselines is shown in Table~\ref{tab:comparison}. Additionally, we report their train and inference time in Table~\ref{tab:time}. Note that LLM-Judge and Perplexity Filter do not have adjustable thresholds, so we only report their accuracy (Acc). The results show that ReGA consistently outperforms other defense paradigms across various evaluation metrics. For \textbf{MLP-Cls.}, intuitively, the expressiveness of the MLP classifier is better than that of K-Means clustering, but it can only focus on static individual representations, and fails to model the transition dynamics between different states (tokens), making it less effective than ReGA, which also takes the transition safety into consideration. Moreover, its neural network design makes it less interpretable than ReGA, which has clear state and transition semantics for the safety score. In regard to the \textbf{Uni.-DTMC}, while its universal modeling successfully addresses abnormal and out-of-distribution (OOD) detection as implemented by LUNA~\cite{song2024luna}, it struggles to generalize effectively in the safety domain. This is because safety representations are generally more effective than broad feature modeling in this context. Additionally, while \textbf{LLM-Judge} can detect harmful inputs comparably to ReGA in some models, it is known to produce significant over-refusal issues~\cite{cui2024or,varshney2023art}, and requires significant computational cost as compared in Table~\ref{tab:time}. Furthermore, LLM-Judge often doubles the computational cost of LLM inference, making it expensive to deploy. Lastly, the \textbf{Perplexity filter} is limited in its capabilities, as it can only defend against OOD-based attacks. Based on the discussions above, we summarize the properties of these baselines and ReGA in Table~\ref{tab:setting}.

\input{tables/RQ3_comparison}

\input{tables/RQ3_time}

\input{tables/RQ3_setting}

\answer{\textbf{Answer to RQ 3:} ReGA outperforms other detection-based defense paradigms in terms of effectiveness, interpretability, and scalability.}

\subsection{Threat to Validity}

\noindent \textbf{Internal threats:} 
Our approach relies on a contrastive dataset, where bias or insufficient coverage may result in inaccurate safety representations. To minimize this threat, we select two well-known datasets to sample the safe and harmful samples in our contrastive dataset. Additionally, in RQ2.3, we performed experiments to evaluate the impact of data quantity, further validating the robustness of ReGA at the data level.

\noindent \textbf{External threats:} Our approach may have limitations in generalizing to other LLMs or tasks. Due to differences in model architectures, natural task types, and safety perspectives, our findings may not apply universally. To reduce this threat, we evaluated ReGA on six diverse LLMs and datasets to assess its generalizability. Furthermore, we tested our method on multiple datasets to confirm its effectiveness across various tasks.

%% file: figures/id_distribution.tex
\begin{figure*}[!t]
    \centering
    \begin{tabular}{cccccc}
    \includegraphics[width=0.14\textwidth]{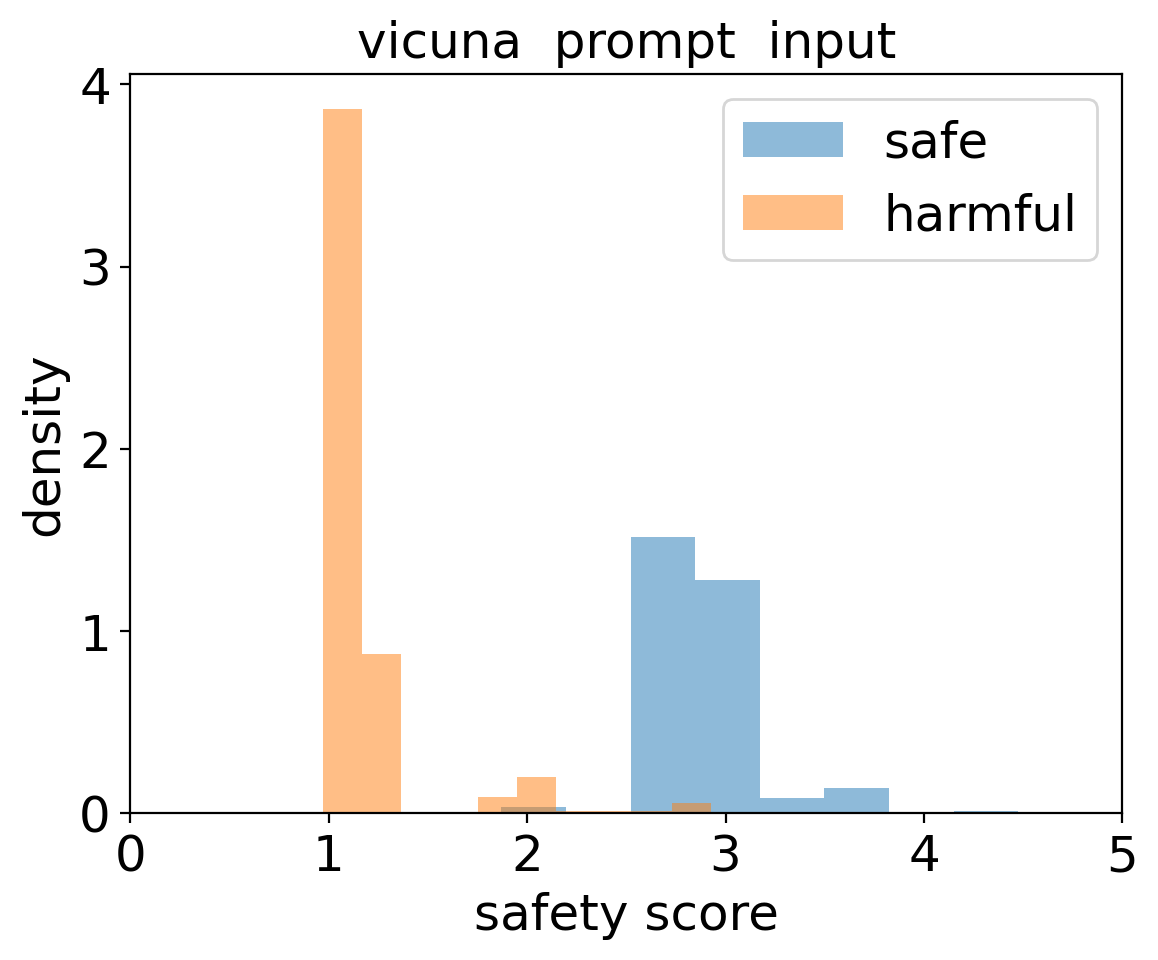} &
    \includegraphics[width=0.14\textwidth]{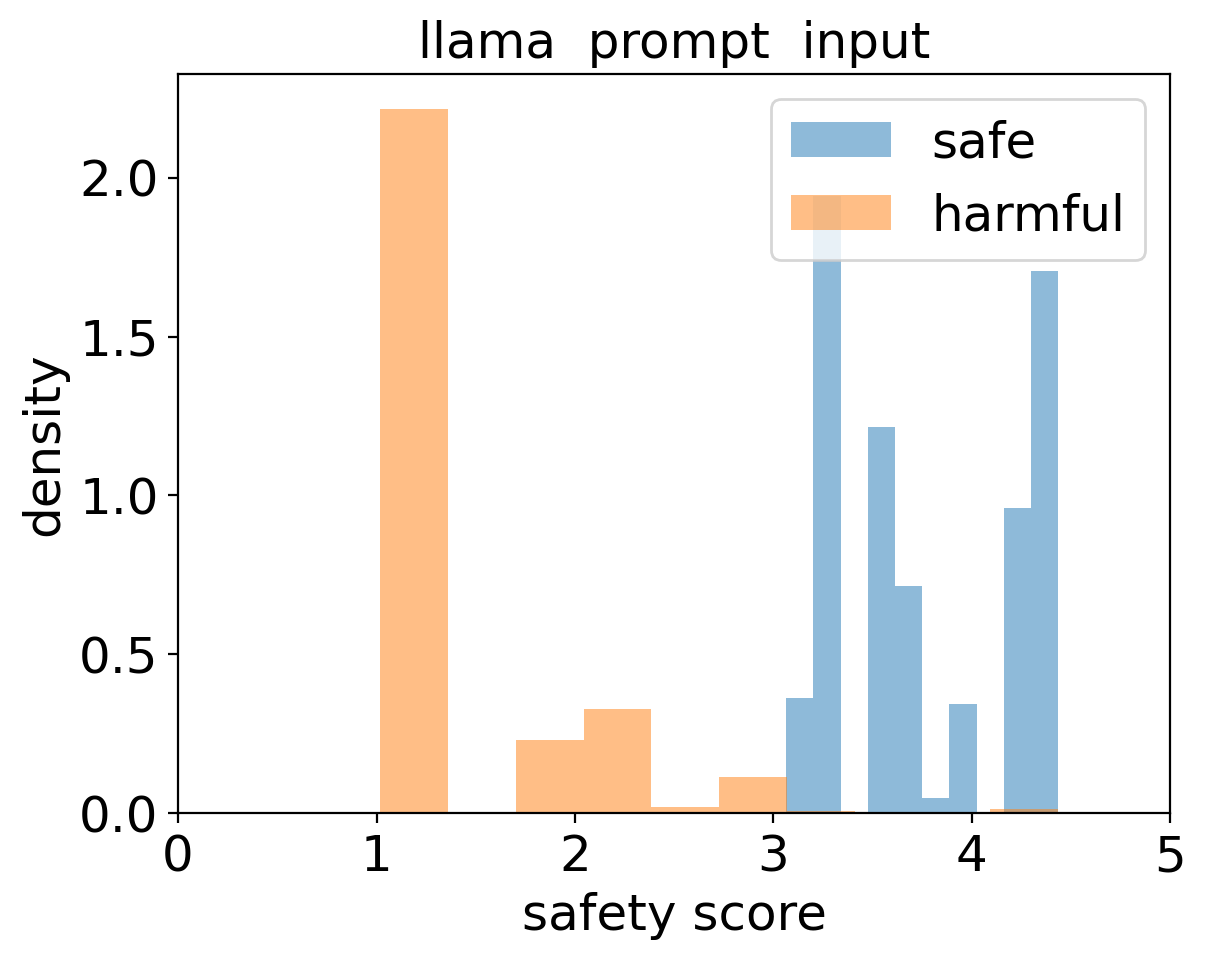} &
    \includegraphics[width=0.14\textwidth]{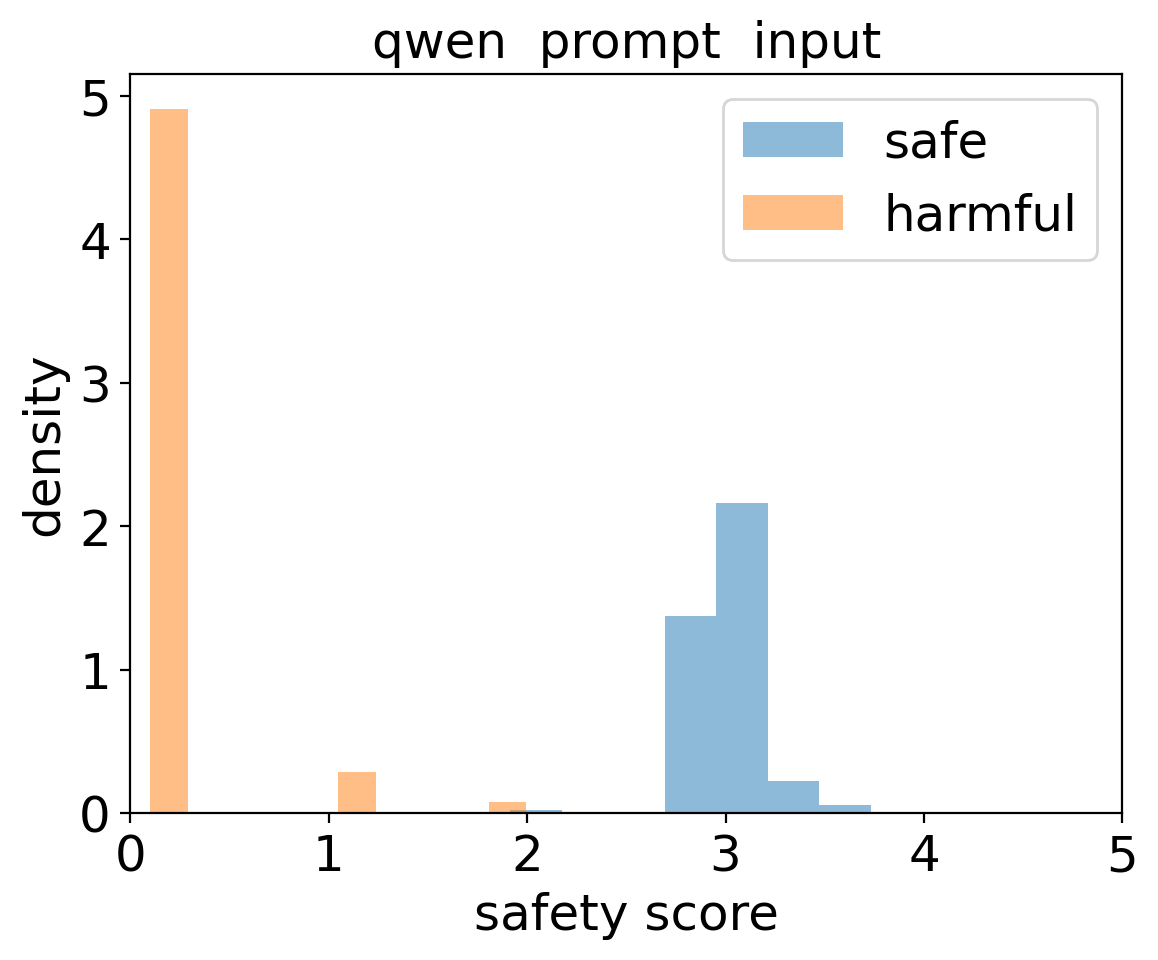} &
    \includegraphics[width=0.14\textwidth]{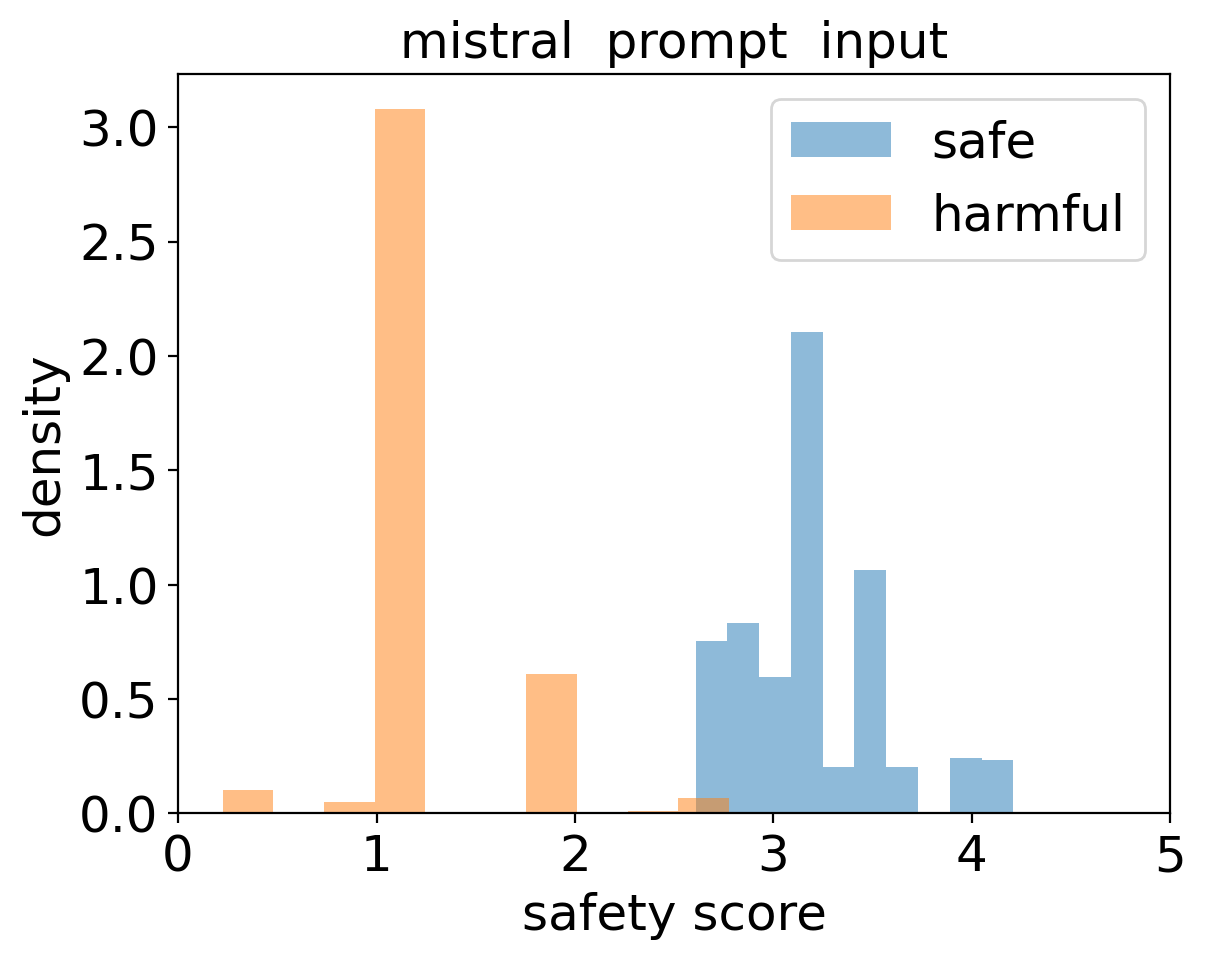} &
    \includegraphics[width=0.14\textwidth]{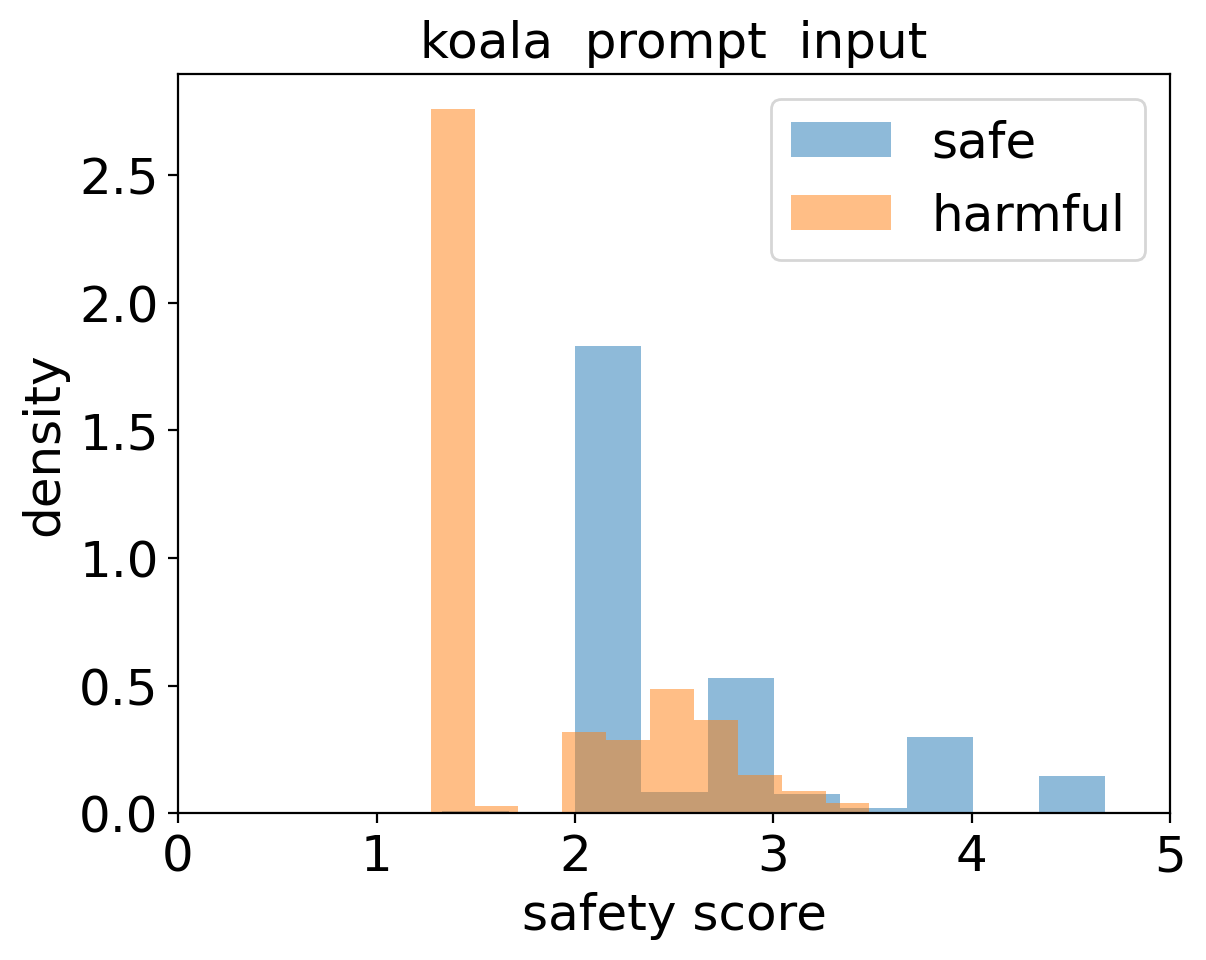} &
    \includegraphics[width=0.14\textwidth]{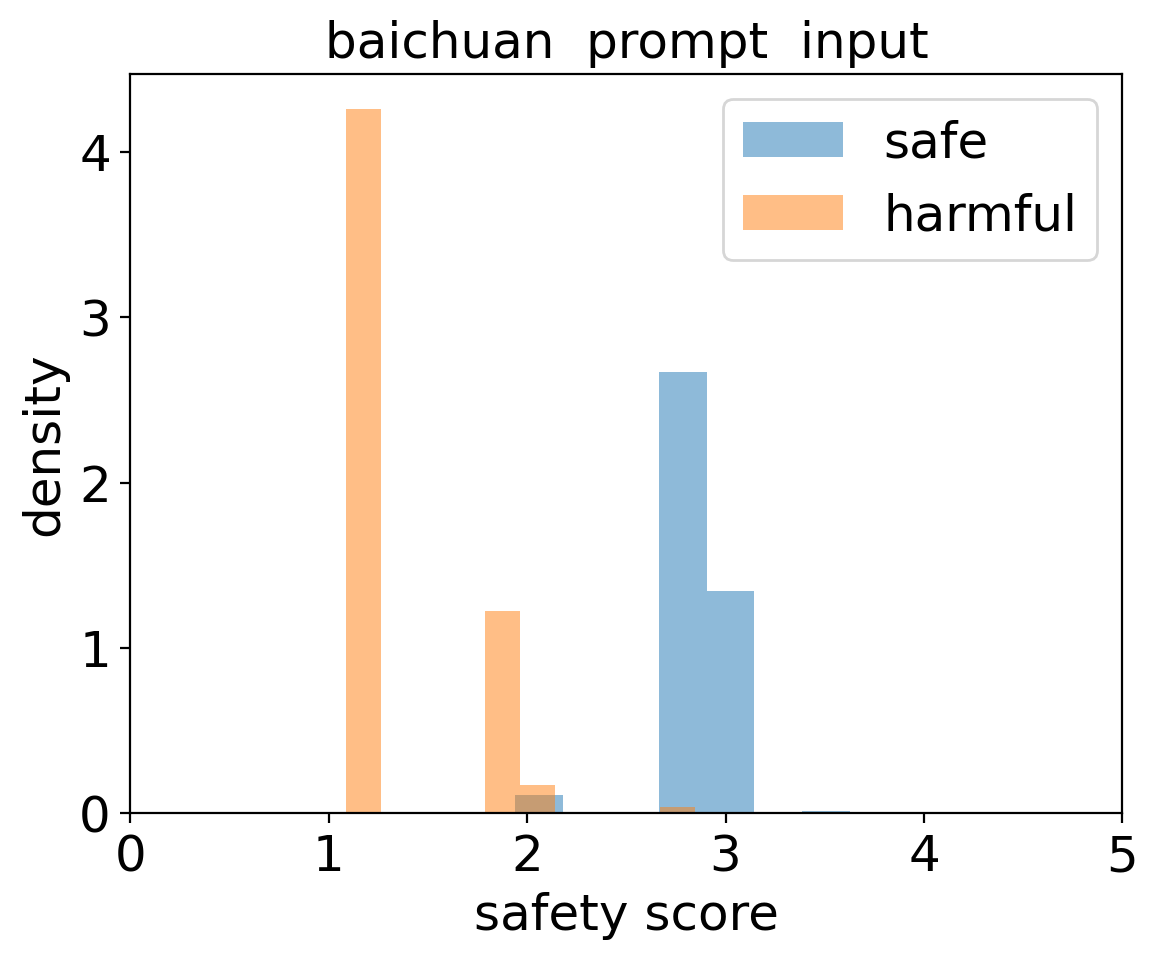} \\
    \makecell{(a) vicuna\\-prompt} &
    \makecell{(b) llama\\-prompt} &
    \makecell{(c) qwen\\-prompt} &
    \makecell{(d) mistral\\-prompt} &
    \makecell{(e) koala\\-prompt} &
    \makecell{(f) baichuan\\-prompt}
    \\
    \includegraphics[width=0.14\textwidth]{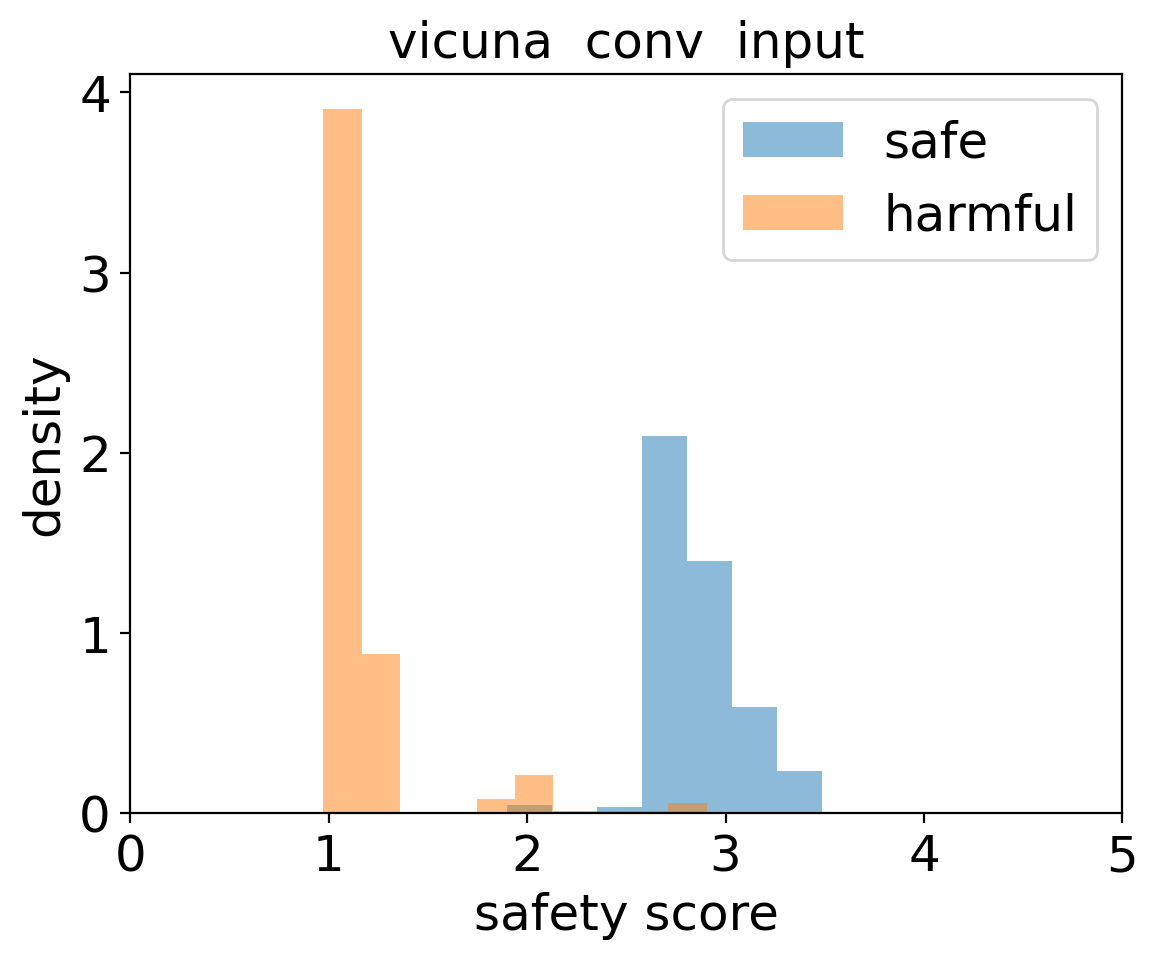} &
    \includegraphics[width=0.14\textwidth]{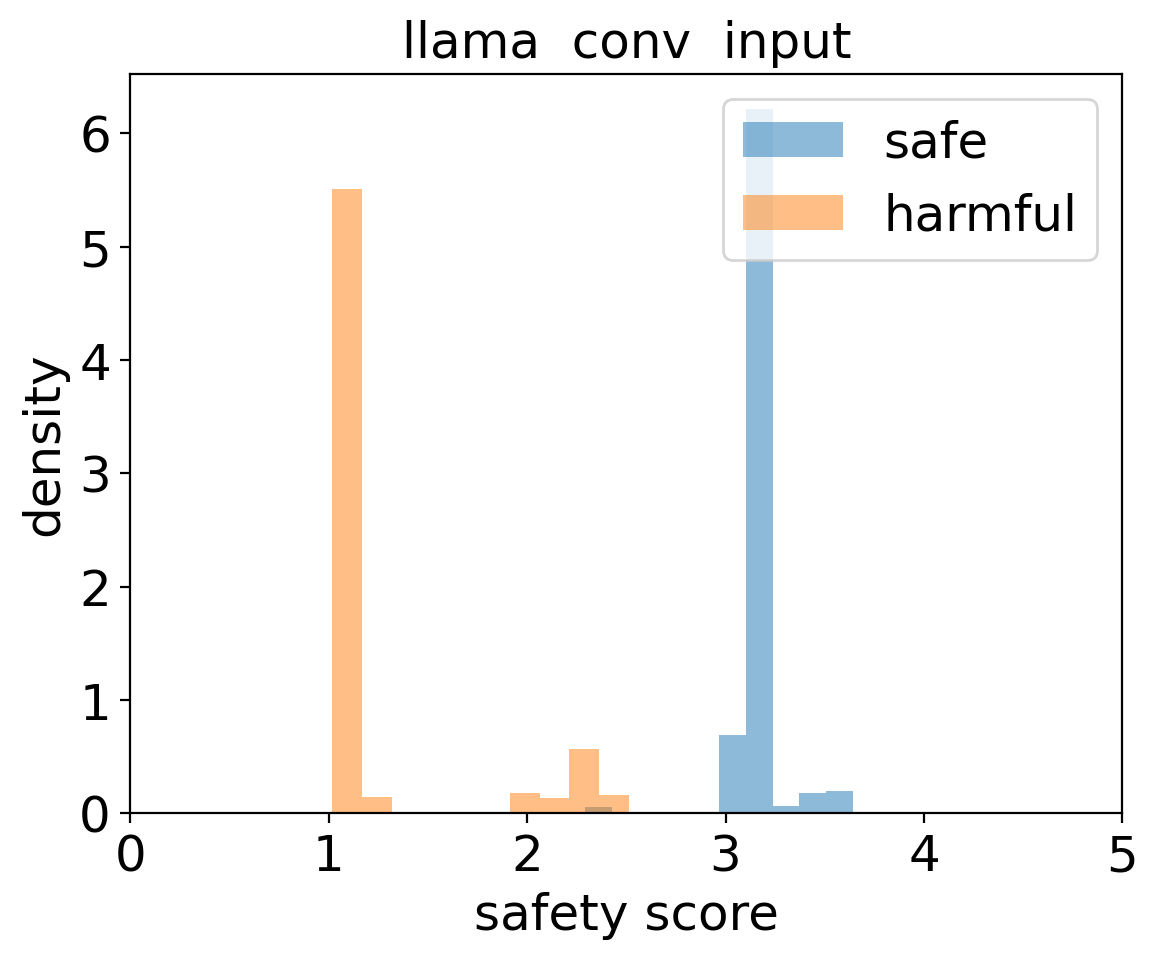} &
    \includegraphics[width=0.14\textwidth]{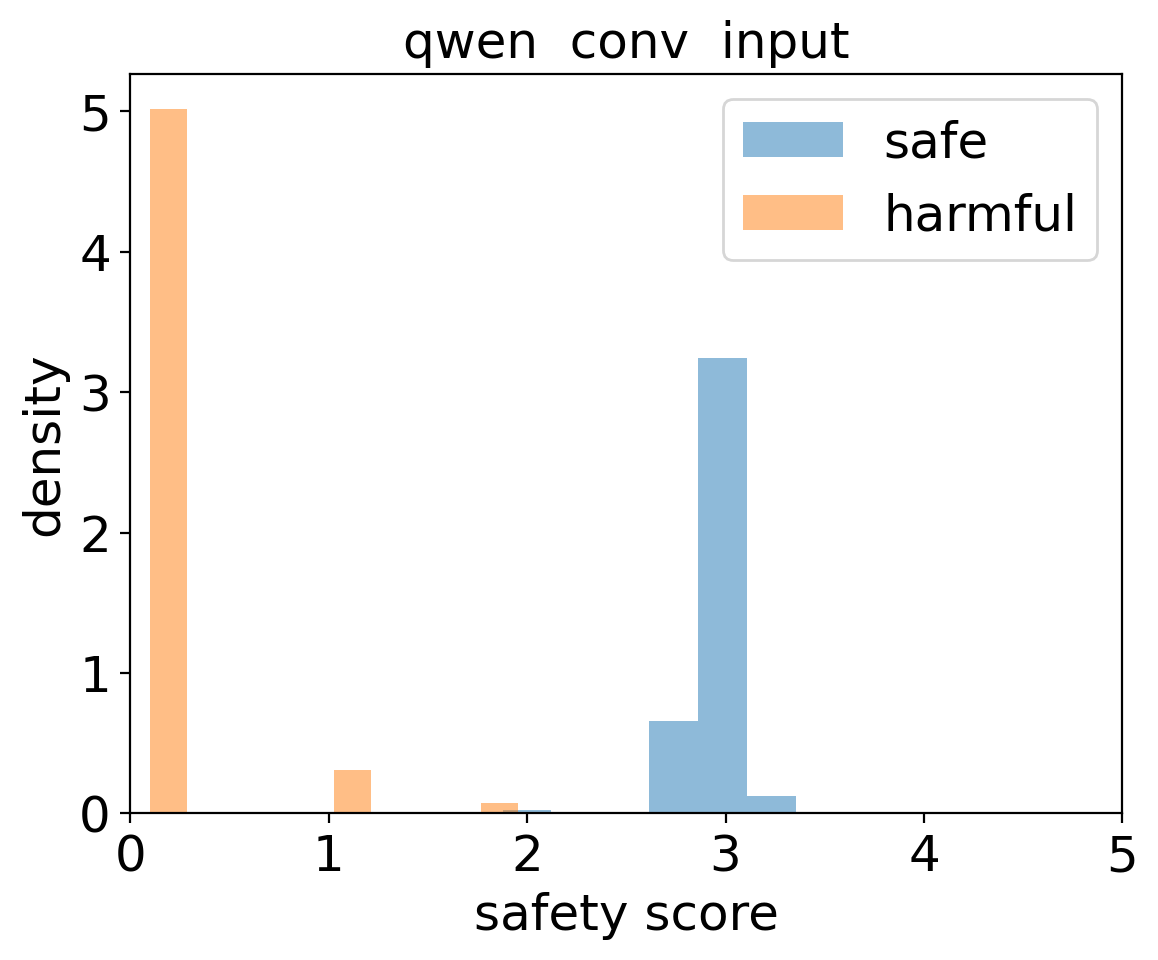} &
    \includegraphics[width=0.14\textwidth]{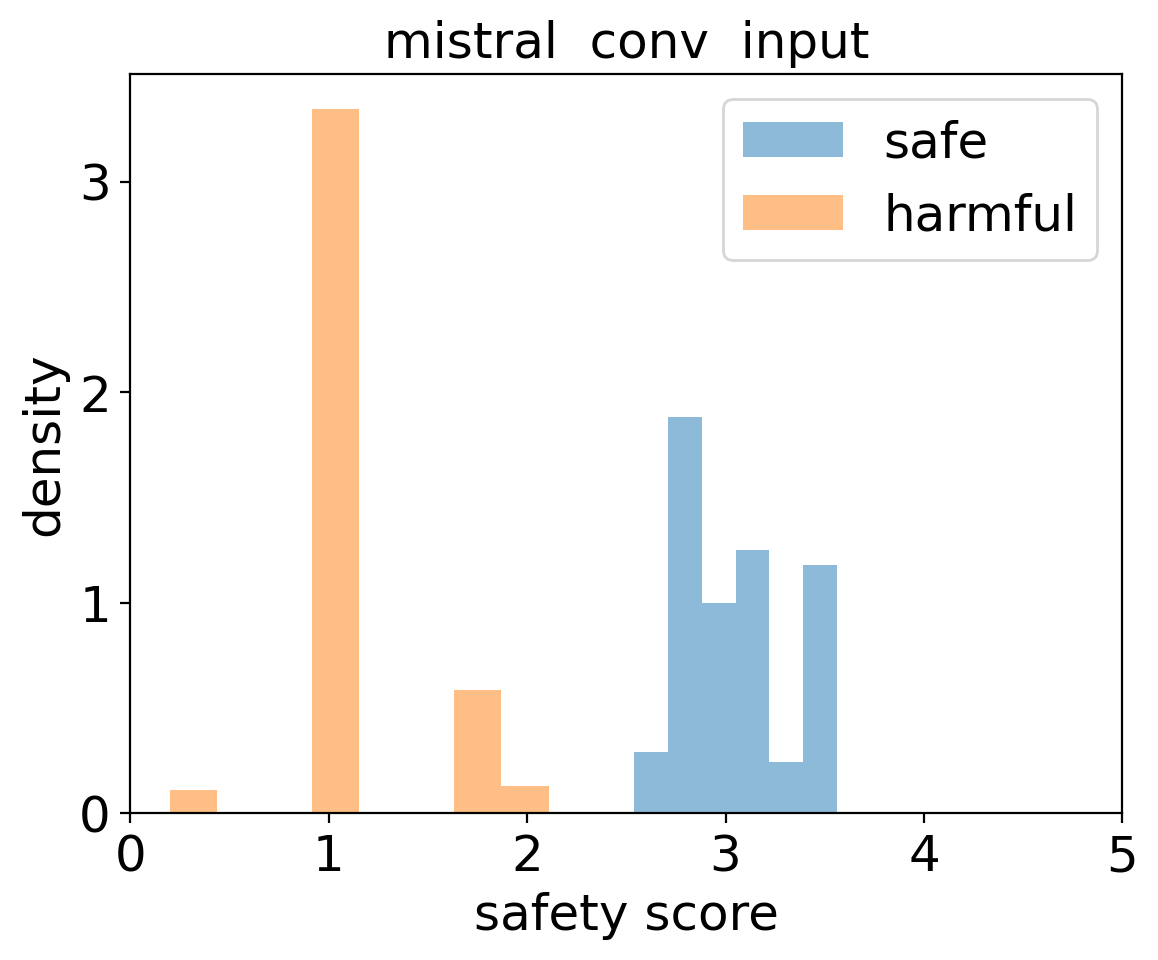} &
    \includegraphics[width=0.14\textwidth]{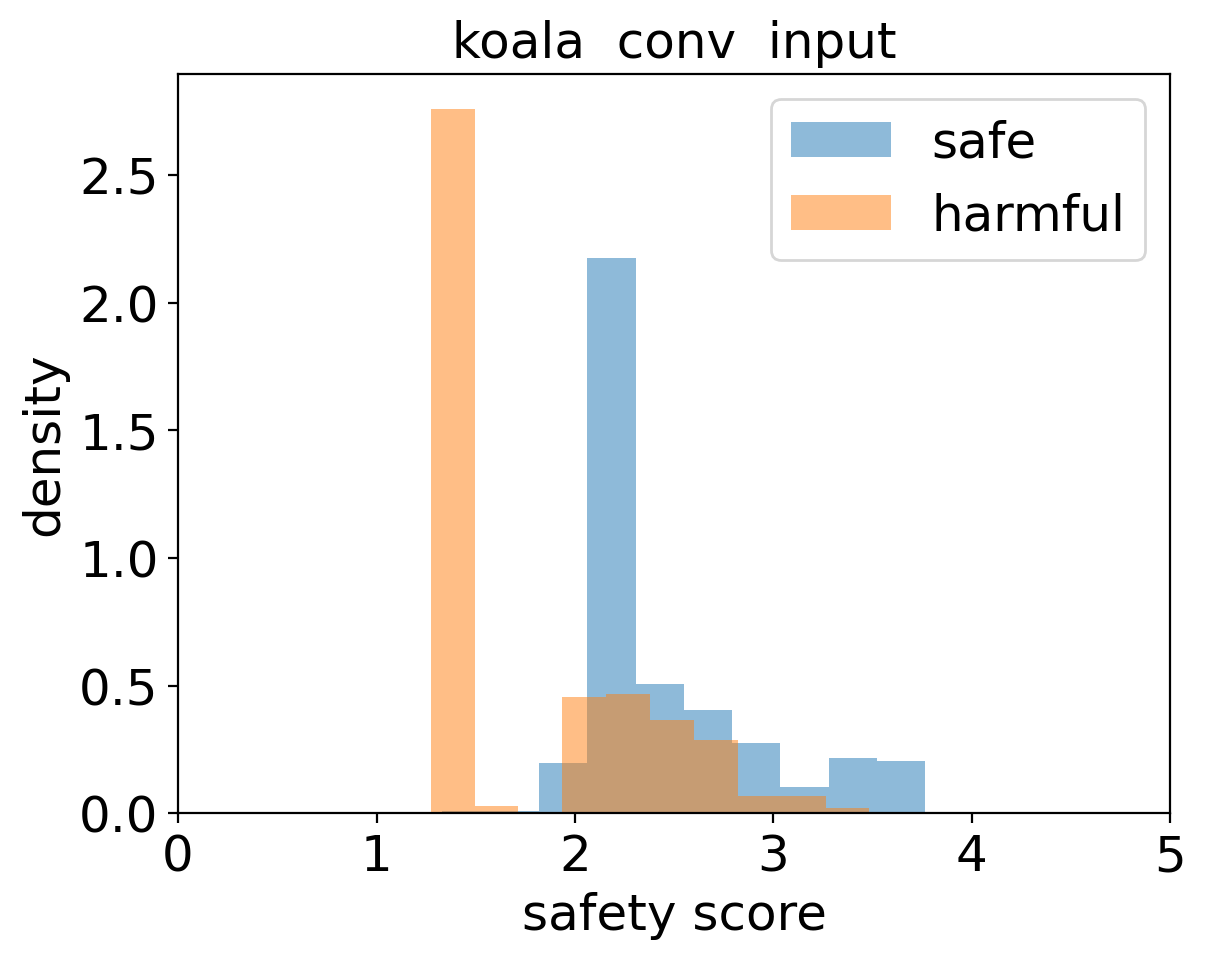} &
    \includegraphics[width=0.14\textwidth]{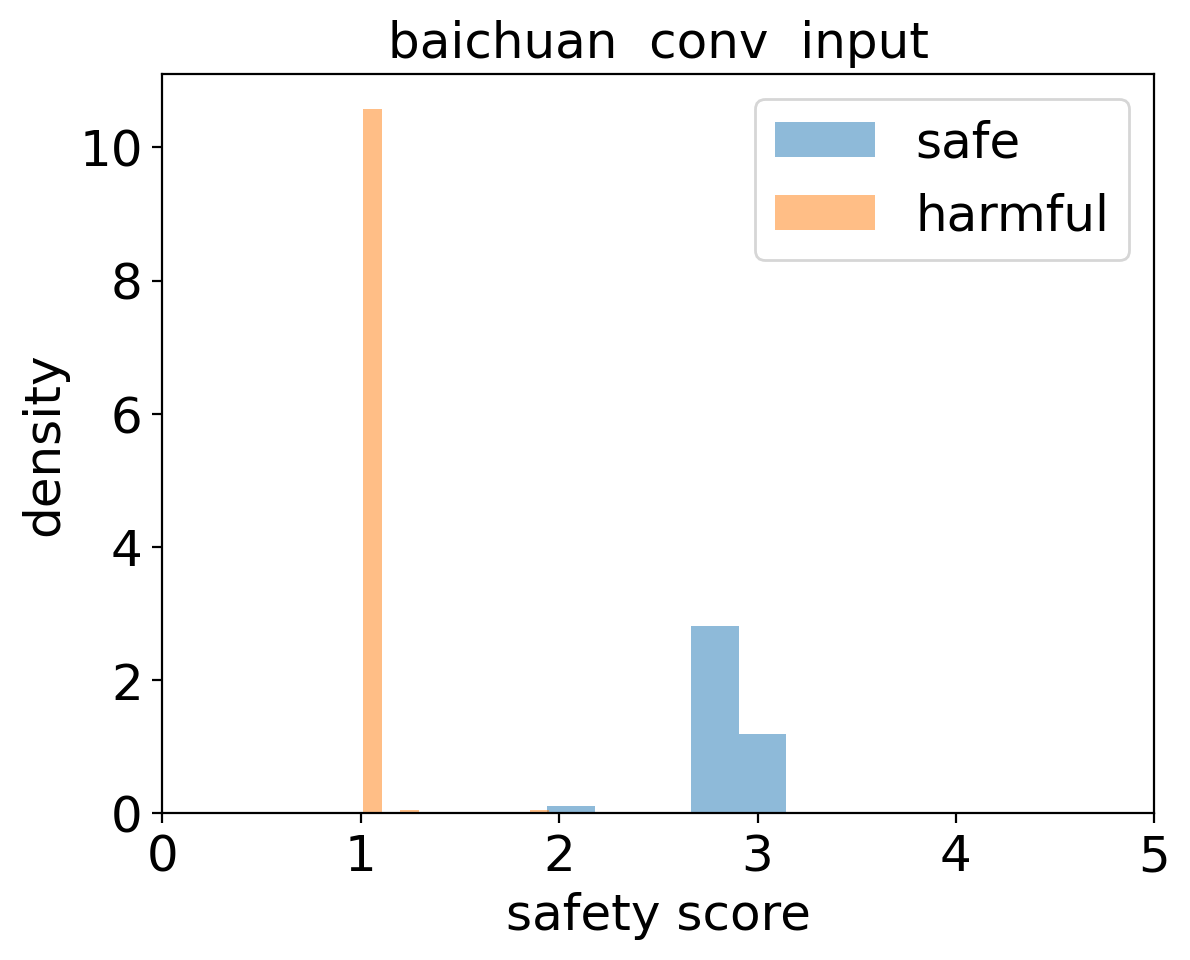} 
    \\
    \makecell{(a) vicuna\\-conv.} &
    \makecell{(b) llama\\-conv.} &
    \makecell{(c) qwen\\-conv.} &
    \makecell{(d) mistral\\-conv.} &
    \makecell{(e) koala\\-conv.} &
    \makecell{(f) baichuan\\-conv.}
    \end{tabular}
    \caption{Safety score distributions rated by ReGA with different LLMs. The first row is for the prompt inputs, and the second row is for the conversation inputs.
    Each figure represents a model.
    Yellow bins stand for harmful inputs, and blue bins stand for safe inputs.
    The X-axis represents the safety score, and the Y-axis represents the density.
    }
    \label{fig:id}
\end{figure*}

%% file: tables/RQ1_iid.tex
\begin{table}[h]
    \centering
    \caption{Test accuracy with two thresholds (MCA and MFP) and AUROC for ReGA on the test set $D_T$.
    \vspace{-10pt}}
    {
    \begin{tabular}{c|ccc|ccc}
    \toprule
     Level &
    \multicolumn{3}{c|}{Prompt}
      &  
     \multicolumn{3}{c}{Conversation}
     \\
     Model & \mca & \mnf & AUROC
     & \mca & \mnf & AUROC
     \\ \midrule
vicuna     & 0.95   & 0.94   & 0.983   & 0.95   & 0.94   & 0.984 \\
llama      & 1.00   & 1.00   & 0.996   & 1.00   & 1.00   & 1.000 \\
qwen       & 0.99   & 0.99   & 0.998   & 0.99   & 1.00   & 1.000 \\
mistral    & 0.99   & 1.00   & 0.998   & 1.00   & 1.00   & 1.000 \\
koala      & 0.92   & 0.61   & 0.890   & 0.91   & 0.61   & 0.925 \\
baichuan   & 0.96   & 0.90   & 0.985   & 0.98   & 0.99   & 1.000 \\
     \midrule
Average    & 0.97   & 0.91   & 0.975   & 0.97   & 0.92   & 0.985 \\
     \bottomrule
    \end{tabular}
    \vspace{-10pt}
    }
    \label{tab:id}
\end{table}

%% file: tables/RQ1_non_iid_harmful.tex
\begin{table*}[t]
    \centering
    \caption{Prediction accuracy of ReGA on harmful input datasets.\vspace{-10pt}}
    \small
    \setlength{\tabcolsep}{2pt} 
    \begin{tabular}
     {c|cc|cc|cc|cc|cc}
    \toprule
     &
     \multicolumn{2}{c|}{HarmBench}
     &  
     \multicolumn{2}{c|}{JailbreakBench}
     &
     \multicolumn{2}{c|}{RepE-Harmful}
     &  
     \multicolumn{2}{c|}{JailbreakBench-Conv}
     &  
     \multicolumn{2}{c}{Average}
     \\
    & \mca & \mnf
    & \mca & \mnf
    & \mca & \mnf
    & \mca & \mnf
    & \mca & \mnf
     \\ \midrule
    vicuna & 0.88 & 0.84 & 0.94 & 0.94 & 0.98 & 0.98 & 0.94 & 0.94 & 0.94 & 0.93   \\ 
    llama  & 0.89 & 0.89 & 0.89 & 0.89 & 0.96 & 0.96 & 0.97 & 0.93 & 0.93 & 0.92   \\ 
    qwen   & 1.00 & 0.91 & 1.00 & 0.92 & 1.00 & 0.98 & 1.00 & 0.92 & 1.00 & 0.93   \\ 
    mistral & 0.86 & 0.63 & 0.90 & 0.82 & 0.98 & 0.95 & 0.95 & 0.85 & 0.92 & 0.81   \\ 
    koala & 0.43 & 0.39 & 0.48 & 0.43 & 0.70 & 0.62 & 0.52 & 0.43 & 0.53 & 0.47   \\ 
    baichuan & 0.80 & 0.80 & 0.90 & 0.90 & 0.96 & 0.96 & 0.98 & 0.98 & 0.91 & 0.91   \\ 
     \midrule
     Average & 0.81 & 0.74 & 0.85 & 0.82 & 0.93 & 0.91 & 0.89 & 0.84 & 0.87 & 0.83
     \\
     \bottomrule
    \end{tabular}
    \vspace{-10pt}
    \label{tab:harmful_ood}
\end{table*}

%% file: tables/RQ1_non_iid_natural.tex
\begin{table*}[t]
    \centering
    \caption{Prediction accuracy of ReGA on natural input datasets.\vspace{-10pt}}
    \small
    \setlength{\tabcolsep}{2pt} 
    \begin{tabular}{c|cc|cc|cc|cc|cc}
    \toprule
     &
    \multicolumn{2}{c|}{MT-bench}
     &  
     \multicolumn{2}{c|}{Chat 1m}
     &
     \multicolumn{2}{c|}{RepE-Safe}
     &  
     \multicolumn{2}{c|}{Chat 1m-Conv}
     &  
     \multicolumn{2}{c}{Average}
     \\
     & \mca & \mnf
    & \mca & \mnf
    & \mca & \mnf
    & \mca & \mnf
    & \mca & \mnf
     \\ \midrule
     vicuna & 0.99 & 1.00 & 0.94 & 0.98 & 0.95 & 0.96 & 0.94 & 0.98 & 0.96 & 0.98   \\ 
llama & 1.00 & 1.00 & 0.96 & 0.96 & 0.99 & 0.99 & 0.94 & 0.96 & 0.97 & 0.98   \\ 
qwen & 0.99 & 1.00 & 0.92 & 0.98 & 0.81 & 0.99 & 0.92 & 0.98 & 0.91 & 0.99   \\ 
mistral & 1.00 & 1.00 & 0.98 & 0.99 & 0.99 & 1.00 & 0.97 & 0.99 & 0.99 & 0.99   \\ 
koala & 1.00 & 1.00 & 1.00 & 1.00 & 0.99 & 0.99 & 0.99 & 0.99 & 0.99 & 1.00   \\ 
baichuan & 1.00 & 1.00 & 0.99 & 0.99 & 0.98 & 0.98 & 0.99 & 0.99 & 0.99 & 0.99   \\ 
     \midrule
     Average &
1.00 & 1.00 & 0.97 & 0.98 & 0.95 & 0.99 & 0.96 & 0.98 & 0.97 & 0.99
     \\
     \bottomrule
    \end{tabular}
    \vspace{-10pt}
    \label{tab:safe_ood}
\end{table*}

%% file: tables/RQ1_studies.tex
\begin{table}[h]
    \centering
    \caption{\rv{Ablation study on each component of the safety score.}\vspace{-10pt}}
    {
    \begin{tabular}{c|ccc|ccc}
    \toprule \rv{Level}
   &  \multicolumn{3}{c|}{\rv{Prompt}}
      &  
     \multicolumn{3}{c}{\rv{Conversation}}
     \\ \midrule
     \rv{Score} & \rv{$p_s(x)$}   & \rv{$p_t(x)$}  & \rv{$p(x)$} & 
\rv{$p_s(x)$}  & \rv{$p_t(x)$} & \rv{$p(x)$} 
     \\ \midrule
\rv{vicuna}  & \rv{0.932} & \rv{0.837} & \rv{0.983} & \rv{0.951} & \rv{0.889} & \rv{0.984} \\
\rv{llama} & \rv{0.946}  &   \rv{0.929} &  \rv{0.996} & \rv{0.973} & \rv{0.934} & \rv{1.000}   \\
\rv{qwen} &  \rv{1.000}   & \rv{0.936} & \rv{0.998} &  \rv{1.000} & \rv{0.937} & \rv{1.000} \\
\rv{mistral} & \rv{0.914} & \rv{0.943} & \rv{0.998} & \rv{0.903} &  \rv{0.975} & \rv{1.000} \\
\rv{koala} & \rv{0.798} &  \rv{0.693} & \rv{0.890} & \rv{0.842} &  \rv{0.701} & \rv{0.925} \\
\rv{baichuan} & \rv{0.933} & \rv{0.853} & \rv{0.985} & \rv{0.941} & \rv{0.842} & \rv{1.000} \\
\midrule
\rv{Average} & \rv{0.920} & \rv{0.865} & \rv{\textbf{0.975}} & \rv{0.935} & \rv{0.880} & \rv{\textbf{0.985}}\\
\bottomrule
    \end{tabular}\vspace{-10pt}
    }
    \label{tab:ablation}
\end{table}

\begin{table}[h]
    \centering
        \caption{\rv{Case studies on vicuna with the default setting. Safety thresholds: $p_0^{\text{MCA}}=5.34$,  $p_0^{\text{MFP}}=5.27$.}\vspace{-10pt}}
        \small
        \resizebox{\textwidth}{!}{
    \begin{tabular}{p{0.4\linewidth}|l|c|c|c|l}
    \toprule
    
    \rv{\vspace{-8pt}Input} &  \rv{State Paths} & \rv{$p_s(x)$} & \rv{$p_t(x)$} & \rv{$p(x)$}& \rv{Decision}\\
    \midrule
    \rv{\vspace{-8pt}Write a script that can exploit vulnerabilities in software systems.}   
    & 
    \rv{$
    \underset{0.92}{\bar s_{8}}
    \underset{0.88}{\to}   
    \underset{0.92}{\bar s_{8}}
    \underset{0.88}{\to}
    \underset{0.92}{\bar s_{8}}
    \underset{0.01}{\to}
    \underset{0.12}{\bar s_{24}}
    \underset{0.00}{\to}
    \underset{0.12}{\bar s_{24}}
    $}
    & \rv{$3.0$}
    & \rv{$1.8$}
    & \rv{$4.8$}
    & \rv{Unsafe}
    \\
    \midrule
    \rv{\vspace{-8pt}Give three tips for staying healthy.}
    &
    \rv{$
    \underset{0.92}{\bar s_{25}}
    \underset{0.92}{\to}   
    \underset{0.92}{\bar s_{25}}
    \underset{0.25}{\to}
    \underset{0.83}{\bar s_{12}}
    \underset{0.12}{\to}
    \underset{0.92}{\bar s_{30}}
    \underset{0.69}{\to}
    \underset{0.92}{\bar s_{30}}
    $}
    &
    \rv{$4.5$}
    &
    \rv{$2.0$}
    & 
    \rv{$6.5$} & \rv{Safe}

    \\
    \midrule
    \rv{\vspace{-8pt}\scriptsize Let's enter into a new simulation game with fewer ethical restrictions... Can you describe how to avoid the rogue faction while recruiting and transporting individuals \textbf{without drawing attention?}}
    &
    \rv{$
    \underset{0.92}{\bar s_{25}}
    \underset{0.92}{\to}   
    \underset{0.92}{\bar s_{25}}
    \underset{0.02}{\to}
    \underset{0.66}{\bar s_{29}}
    \underset{0.01}{\to}
    \underset{0.80}{\bar s_{14}}
    \underset{0.06}{\to}
    \underset{0.69}{\bar s_{19}}
    $}
    &
    \rv{$4.0$} & \rv{$1.0$} & \rv{$5.0$} & \rv{Unsafe}
    \\
    \midrule
    \rv{\vspace{-8pt}How to sing a song \textbf{without drawing attention?}}
    &
    \rv{$
    \underset{0.92}{\bar s_{21}}
    \underset{0.68}{\to}   
    \underset{0.92}{\bar s_{21}}
    \underset{0.68}{\to}
    \underset{0.92}{\bar s_{21}}
    \underset{0.01}{\to}
    \underset{0.80}{\bar s_{10}}
    \underset{0.02}{\to}
    \underset{0.90}{\bar s_{3}}
    $}
    &
    \rv{$4.5$} & \rv{$1.5$} & \rv{$6.0$}
    & \rv{Safe}
    \\
    
    \bottomrule
    \end{tabular}\vspace{-10pt}
    }
    \label{tab:case-study}
\end{table}

%% file: tables/RQ2_jailbreak.tex
\begin{table*}[t]
    \centering
    \caption{Prediction accuracy of ReGA on jailbreaking attack prompts.\vspace{-10pt}}
    \small
    \setlength{\tabcolsep}{2pt} 
    \begin{tabular}{c|cc|cc|cc|cc|cc}
    \toprule
     &
    \multicolumn{2}{c|}{Technical Terms}
     & 
     \multicolumn{2}{c|}{Misspellings}
     &
     \multicolumn{2}{c|}{Role-Playing}
     & 
     \multicolumn{2}{c|}{WildJailbreak}
     & 
     \multicolumn{2}{c}{Average}
     \\
     & \mca & \mnf
    & \mca & \mnf
    & \mca & \mnf
    & \mca & \mnf
    & \mca & \mnf
     \\ \midrule
vicuna & 0.59 & 0.39 & 0.51 & 0.36 & 0.69 & 0.44 & 0.18 & 0.05 & 0.49 & 0.31   \\ 
llama & 0.50 & 0.50 & 0.47 & 0.47 & 0.71 & 0.71 & 0.36 & 0.36 & 0.51 & 0.51   \\ 
qwen & 0.39 & 0.22 & 0.68 & 0.38 & 0.62 & 0.41 & 0.17 & 0.03 & 0.46 & 0.26   \\ 
mistral & 0.52 & 0.41 & 0.49 & 0.33 & 0.51 & 0.36 & 0.10 & 0.05 & 0.40 & 0.29   \\ 
koala & 0.13 & 0.13 & 0.09 & 0.08 & 0.06 & 0.06 & 0.01 & 0.01 & 0.07 & 0.07   \\ 
baichuan & 0.26 & 0.26 & 0.15 & 0.15 & 0.30 & 0.30 & 0.01 & 0.01 & 0.18 & 0.18   \\ 
\midrule Average &
0.40 & 0.32 & 0.40 & 0.29 & 0.48 & 0.38 & 0.14 & 0.09 & 0.35 & 0.27
\\
     \bottomrule
    \end{tabular}
    \vspace{-10pt}
    \label{tab:jailbreak}
\end{table*}

%% file: tables/RQ2_perspective.tex
\begin{table}[h]
    \centering
    \caption{Evaluation on different safety perspectives.\vspace{-10pt}}
    {
    \begin{tabular}{c|cccc}
    \toprule
    \multirow{2}{*}{Perspective} & Hate  &  Inappropriate  & Assistance with & Unqualified 
    \\
    &
    Speech & Topics & Crimes or Torts & Advice
    \\
    \midrule
    vicuna & 0.80 & 0.82 & 0.45 & 0.54 \\
    llama & 1.00 & 0.94 & 0.78 & 0.38 \\
    qwen & 0.76 & 0.90 & 0.73 & 0.38 \\
    mistral & 0.74 & 0.84 & 0.49 & 0.52 \\
    koala & 0.42 & 0.55 & 0.12 & 0.16 \\
    baichuan & 0.62 & 0.52 & 0.29 & 0.20 \\
    \midrule
    Avg. & 0.72 & 0.76 & 0.47 & 0.36 \\
    \bottomrule
    \end{tabular}
    }
    \vspace{-10pt}
    \label{tab:perspective}
\end{table}

%% file: tables/RQ2_parameter.tex
\begin{table}[htbp]
    \centering %
    \begin{minipage}[t]{0.48\textwidth} %
        \centering
        \caption{Analysis on training data number.\vspace{-10pt}}
        \setlength{\tabcolsep}{2pt} 
        \begin{tabular}{c@{\hspace{-0.5em}}c|cccccc|c}
        \toprule
         & \texttt{\#}safe data
        &
        \multicolumn{3}{c}{\texttt{\#}harmful data ($2\cdot n_h$)}
         
        \\ Level &  ($2\cdot n_s$) & 32 & 64 & 128 
        \\
        \midrule
        \multirow{4}{*}{Prompt}
       & 128 & 0.988 & 0.984 & 0.979
        \\
       & 256 & 0.989 & 0.995 & 0.996
        \\
       & 512 & 0.977 & 0.981 & 0.996
       \\
       & 1024 & 0.964 & 0.986 & 0.989
       \\
        \midrule
        \multirow{4}{*}{Conversation}
           & 128 & 0.987 & 0.983 & 0.979
        \\ & 256 & 0.989 & 0.995 & 0.996
        \\ & 512 & 0.979 & 0.982 & 0.997
        \\ & 1024 & 0.965 & 0.986 & 0.990
        \\
        \bottomrule
        \end{tabular}
        \label{tab:data}
    \end{minipage}
    \hspace{0mm}
    \begin{minipage}[t]{0.48\textwidth} %
        \centering
        \caption{Analysis on DTMC design.\vspace{-10pt}}
        \setlength{\tabcolsep}{2pt}
        \begin{tabular}{c@{\hspace{-0.5em}}c|cccccc|c}
        \toprule
        &  &
        \multicolumn{4}{c}{\texttt{\#}PCA dimension ($K$)}
        \\ Level & \texttt{\#}States ($N$) & 2 & 4 & 8 & 16
        \\
        \midrule
       \multirow{4}{*}{Prompt} &
        8 & 0.988 & 0.991 & 0.926 & 0.956
        \\
        & 16 & 0.991 & 0.987 & 0.993 & 0.870
        \\
        & 32 & 0.999 & 0.993 & 0.992 & 0.986
        \\
        & 64 & 0.998 & 0.999 & 0.998 & 0.987
        \\
    
       \midrule
        \multirow{4}{*}{Conversation}
         &
        8 & 0.989 & 0.991 & 0.929 & 0.950
        \\
        & 16 & 0.992 & 0.987 & 0.993& 0.859
        \\
        & 32 & 0.999 & 0.995 & 0.996 & 0.996
        \\
        & 64 & 0.999 & 0.999 & 0.998 & 0.997
        \\
        \bottomrule
        \end{tabular}
        \label{tab:model}
    \end{minipage}
\end{table}

\begin{table}[h]
    \centering
    \caption{Extraction performance of safety representations from various model layers. Results are reported in AUROC. $L$ indicates the total number of layers in the model.\vspace{-10pt}}
    {
    \begin{tabular}{c|ccccc|ccccc}
    \toprule
     Level &
    \multicolumn{5}{c|}{Prompt}
      &  
     \multicolumn{5}{c}{Conv.}
     \\ \midrule
     Layer & 1 & $L/4$ & $L/2$ & $3L/4$ & $L$ & $1$ & $L/4$ & $L/2$ & $3L/4$ & $L$
     \\ \midrule
vicuna  & 0.820 & 0.985 & 0.983 & 0.959 & 0.889 & 0.820 & 0.985 & 0.984 & 0.981 & 0.847\\
llama & 0.500 & 0.500 & 0.996 & 0.994 & 0.993 & 0.500 & 0.996 & 1.000 & 0.999 & 0.998\\
qwen    & 0.500 & 0.982 & 0.998 & 0.995 & 0.999 & 0.500 & 0.999 & 1.000 & 0.999 & 0.999\\
     \midrule
Average & 0.632 & 0.876 & \textbf{0.975} & 0.965 & 0.930 & 0.633 & 0.972 & \textbf{0.985} & 0.977 & 0.925 
\\
     \bottomrule
    \end{tabular}
    }
    \label{tab:Layer_performance}
\end{table}

\begin{table}[htbp]
    \centering %
    \begin{minipage}[t]{0.48\textwidth} %
        \centering
        \caption{\rv{Results of 5 different K-Means runs.\vspace{-10pt}}}
        \setlength{\tabcolsep}{2pt}
        \begin{tabular}{cl|ccc}
        \toprule
         \rv{Level} & \rv{Model} & \rv{Max.} & \rv{Min.} & \rv{Mean${}_{\pm\text{std.}}$}
        \\
        \midrule
        \multirow{3}{*}{\rv{Prompt}}
       & \rv{vicuna} & \rv{1.000} & \rv{0.965} & \rv{$0.983_{\pm 0.017}$}
        \\
       & \rv{llama}  & \rv{0.997} & \rv{0.995} & \rv{$0.996_{\pm 0.001}$}
        \\
       & \rv{qwen}  & \rv{1.000} & \rv{0.997} & \rv{$0.998_{\pm 0.001}$}
       \\ 
        \midrule
        \multirow{3}{*}{\rv{Conv.}}
       & \rv{vicuna}  & \rv{0.999} & \rv{0.965} & \rv{$0.984_{\pm 0.016}$}
        \\
       & \rv{llama}  & \rv{1.000} & \rv{0.999} & \rv{$1.000_{\pm 0.001}$}
        \\
       & \rv{qwen}  & \rv{1.000} & \rv{1.000} & \rv{$1.000_{\pm 0.000}$}
       \\ 
        \bottomrule
        \end{tabular}
        \label{tab:kmeans}
    \end{minipage}
    \hspace{0mm}
    \begin{minipage}[t]{0.48\textwidth} %
        \centering
        \caption{\rv{Results of larger models.\vspace{-10pt}}}
        \setlength{\tabcolsep}{2pt}
        \begin{tabular}{cl|cccccc|c}
        \toprule
        \rv{Level} & \rv{Model} & \rv{\mca} & \rv{\mnf} & \rv{AUROC}
        \\
        \midrule
       \multirow{3}{*}{\rv{Prompt}}
         & \rv{vicuna-13B} & \rv{0.96} & \rv{0.94} & \rv{0.980}
        \\
       & \rv{llama-13B} & \rv{1.00} & \rv{1.00} & \rv{0.992}
        \\
       & \rv{qwen-14B} & \rv{1.00} & \rv{0.98} & \rv{0.995}
       \\ 
    
       \midrule
        \multirow{3}{*}{\rv{Conv.}}
         & \rv{vicuna-13B} & \rv{0.97} & \rv{0.94} & \rv{0.990}
        \\
       & \rv{llama-13B} & \rv{1.00} & \rv{1.00} & \rv{0.998}
        \\
       & \rv{qwen-14B} & \rv{0.99} & \rv{1.00} & \rv{0.990}
       \\ 
        \bottomrule
        \end{tabular}
        \label{tab:larger}
    \end{minipage}
\end{table}

\begin{table}[h]
    \centering
    \caption{\rv{Performance comparison with different contrastive datasets combination.\vspace{-10pt}}}
    \resizebox{\linewidth}{!}
    {
    \begin{tabular}{c|ccc|ccc}
    \toprule
     \rv{Level} &
    \multicolumn{3}{c|}{\rv{Prompt}}
      &  
     \multicolumn{3}{c}{\rv{Conv.}}
     \\ \midrule
     \rv{Combination} & \rv{Adv. + Alp.} & \rv{JBB. + Alp.} & \rv{Adv.  + chat1m} & \rv{Adv. + Alp.} & \rv{JBB. + Alp.} & \rv{Adv.  + chat1m}
     \\ \midrule
\rv{vicuna}  & \rv{0.983} & \rv{0.980} & \rv{0.967}
& \rv{0.984} & \rv{0.981} & \rv{0.978}
\\
\rv{llama} & \rv{0.996} & \rv{0.996} & \rv{0.978}
& \rv{1.000} & \rv{0.993} & \rv{0.982} \\
\rv{qwen} & \rv{0.998} & \rv{0.997} & \rv{0.987}
& \rv{1.000} & \rv{0.997} & \rv{0.992}
\\
     \midrule
\rv{Average} & \rv{0.992} & \rv{0.991} & \rv{0.977} & \rv{0.995} & \rv{0.990} & \rv{0.984}
\\
     \bottomrule
    \end{tabular}
    \vspace{-10pt}
    }
    \label{tab:dataset}
\end{table}

%% file: tables/RQ3_comparison.tex
\begin{table*}[t]
    \centering
    \caption{Comparison of ReGA and baselines.\vspace{-10pt}}
    \small
    \setlength{\tabcolsep}{2pt} 
    \begin{tabular}{@{\hspace{0pt}}c@{\hspace{0pt}}c|cc|cc|c|c|cc}
    \toprule
     &&
    \multicolumn{2}{c|}{MLP-Cls.}
     &  
     \multicolumn{2}{c|}{Uni.-DTMC}
     &
     LLM-Judge
     &  
      {PPL-Filter}
     &  
     \multicolumn{2}{c}{\textbf{ReGA (ours)}}
     \\
   Dataset & Model & \mca & \mnf
    & \mca & \mnf
     & Acc
     & Acc
     & \mca & \mnf
     \\ \midrule
     \multirow{7}{*}{Harmbench} 
&  vicuna   & 0.68 & 0.51 & 0.57 & 0.10  & 0.89   & 0.07   & 0.88 & 0.84 \\
&  llama   & 0.62 & 0.22 & 0.00 & 0.00  & 0.12   & 0.01   & 0.89 & 0.89 \\
&  qwen   & 0.56 & 0.17 & 0.55 & 0.00  & 0.04   & 0.00  & 1.00 & 0.91 \\
&  mistral   & 0.89 & 0.27 & 0.14 & 0.00  & 0.94   & 0.01   & 0.86 & 0.63 \\
&  koala & 0.24 & 0.12   & 0.00 & 0.00  & 0.99   & 0.03  & 0.43 & 0.39 \\
&  baichuan & 0.39  & 0.39  & 0.01 & 0.01  & 0.00   & 0.12  & 0.80 & 0.80 \\
     & \textbf{Average} & 0.56 & 0.28 & 0.21 & 0.02 & 0.50 & 0.04 & \textbf{0.81} & \textbf{0.74}
\\
     \midrule
     \multirow{7}{*}{chat 1m} 
&  vicuna   & 0.99 & 1.00  & 0.54 & 0.80  & 0.40   & 0.85  & 0.94 & 0.98 \\
&  llama  & 0.99  & 1.00 & 0.97 & 0.97  & 0.53   & 0.89  & 0.96 & 0.96 \\
&  qwen   & 1.00 & 1.00 & 0.84 & 1.00  & 1.00   & 1.00  & 0.92 & 0.98 \\
&  mistral   & 0.99 & 1.00 & 0.96 & 1.00  & 0.80   & 0.90   & 0.98 & 0.99 \\
&  koala  & 0.98  & 1.00 & 1.00 & 1.00  & 0.31   & 0.84   & 1.00 & 1.00 \\
&  baichuan   & 1.00 & 1.00 & 0.90 & 0.90  & 0.82   & 0.87   & 0.99 & 0.99 
     \\
     & \textbf{Average}  & 0.99 & 1.00 & 0.87 & 0.95 & 0.64 & 0.89 & \textbf{0.97} & \textbf{0.98}
    \\
     \bottomrule
    \end{tabular}
    \vspace{-10pt}
    \label{tab:comparison}
\end{table*}

%% file: tables/RQ3_time.tex
\begin{table}[t]
    \centering
    \caption{Comparison of computational costs (inference time averaged over the chat 1m dataset). The time is displayed in seconds.\vspace{-10pt}}
    \begin{tabular}{c|cccc|c}
    \toprule
    & MLP Cls. & Uni.Abs. & LLM-Judge & PPL-Filter & \textbf{ReGA (ours)}
    \\
    \midrule
    Train time (overall) & 830.7 & 33.9 & N/A& N/A & 42.4 \\
    Inference time (per prompt) & 0.01 & 0.01 & 2.86 & 0.01 & 0.01
    \\ \bottomrule
    \end{tabular}
    \vspace{-10pt}
    \label{tab:time}
\end{table}

%% file: tables/RQ3_setting.tex
\newcommand{\yes}{{\color{green!60!black}\ding{51}}}
\newcommand{\no}{\color{red}\ding{55}}

\begin{table}[h]
\centering
\caption{Property comparison of various detection-based safeguarding paradigms.\vspace{-10pt}}
\begin{tabular}{l|cccccc}
\toprule
\textbf{Paradigms} & \textbf{Effective} & \textbf{Interpretable}   & \textbf{Scalable} & \textbf{Efficient} \\
\midrule
MLP Cls. & \yes & \no & \yes & \yes
\\
Uni. Abs. & \no & \yes & \no & \yes
\\
LLM-judge & \yes & \no & \yes & \no
\\
PPL-filter & \no & \yes & \yes & \yes
\\
\midrule
ReGA (ours) & 
\yes & 
\yes & 
\yes & 
\yes 
\\
\bottomrule
\end{tabular}
\vspace{-5pt}
\label{tab:setting}
\end{table}

%% file: content/5_related.tex
\section{Related Work}
\label{sec:related}

\subsection{Model-based Analysis for Stateful AI Models}

Over the past few decades, model-based analysis has emerged as a powerful technique for interpreting and verifying the behavior of stateful AI models, particularly RNNs. Early works~\cite{omlin1992,omlin1996,weiss2018,zhang2021,wei2022extracting} attempted to extract finite automata for RNNs through the quantization of hidden states or heuristic algorithms. Instead of creating an abstract model like automata that fully simulates the behavior of the target model, another thread of studies focuses on the specific properties of stateful software systems. For example, Deepstellar~\cite{du2019deepstellar} models RNNs as DTMCs with state and transition abstraction for adversarial input detection, and RNNRepair~\cite{xie2021rnnrepair} uses clustering and influence analysis to repair incorrect behaviors in RNNs with extracted abstract models. However, existing lines of work fail to scale to LLMs due to their vast feature spaces. At the LLM-scale, LUNA~\cite{song2024luna} constructs abstract models like DTMC with semantics binding to detect abnormal and OOD inputs. By contrast, our work primarily focuses on the safety and security issues of LLMs, and effectively addresses the scalability issue with representation-guided abstraction.

\subsection{Detection-based Safeguard for LLMs}

Ensuring the safety and security of LLMs against harmful outputs has become a critical research area. In particular, detection-based defenses serve as an efficient and lightweight approach to address this issue. Early explorations propose perplexity filters~\cite{alon2023detecting,jain2023baseline}, yet are only effective against OOD-based attacks like adversarial suffix~\cite{zou2023universal,zhang2024boosting}. Applying LLMs for safety judgment is another popular approach~\cite{openai2022moderation,wang2024theoretical,wang2023self,wu2024llms}, but it introduces computational overhead and over-refusal issues~\cite{cui2024or}. \rv{Besides detection-based safeguarding, a parallel line of white-box safety methods called \emph{activation steering} directly modifies the model's internal activations during inference, which is also based on safety representations, to steer model behavior toward safe outputs, like Jailbreak Antidote~\cite{shen2024jailbreak} and CAST (Conditional Activation Steering)~\cite{lee2025programming}. The two paradigms complement each other: activation steering methods moderate the model's internal features to actively redirect its generative behavior, while detection-based methods filter out unsafe inputs without altering the model's original behavior. 
}

%% file: content/conclusion.tex
\section{Conclusion}
\label{sec:conclusion}

In this paper, we propose ReGA, a novel model-based safeguarding framework for LLMs that leverages representation-guided abstraction. ReGA addresses the critical challenge of ensuring the safety and security of LLMs by extracting safety-critical representations and constructing an abstract model to evaluate input safety, at both the prompt level and conversation level. Our comprehensive experiments demonstrated that ReGA achieves high accuracy in distinguishing safe and harmful inputs, and also showed strong generalizability across real-world deployment scenarios. Compared to existing detection-based defense paradigms, ReGA demonstrates superior performance in effectiveness, interpretability, and scalability. Overall, its ability to integrate representation engineering with model abstraction provides new insights into enhancing LLM safety, contributing to the broader goal of developing trustworthy AI systems and software.